\documentclass[twocolumn]{aastex631}
\usepackage{graphicx} 
\usepackage{longtable}
\usepackage{natbib}
\usepackage{amsmath, amsfonts}
\usepackage{amssymb}
\usepackage{gensymb}
\usepackage{url}
\usepackage[version=4]{mhchem}

\DeclareUnicodeCharacter{2212}{-}

\begin{document}

\title{Detectability of the Yarkovsky Effect in the Main Belt}
\author{Denise Hung}
\affiliation{Institute for Astronomy, University of Hawai'i, 2680 Woodlawn Drive, Honolulu, HI 96822, USA}

\author{David J. Tholen}
\affiliation{Institute for Astronomy, University of Hawai'i, 2680 Woodlawn Drive, Honolulu, HI 96822, USA}

\author{Davide Farnocchia}
\affiliation{Jet Propulsion Laboratory, California Institute of Technology, Pasadena, CA 91109, USA}

\author{Federica Spoto}
\affiliation{Minor Planet Center--Center for Astrophysics, Harvard \& Smithsonian, 60 Garden Street, MS 15, Cambridge, MA, USA}

\shorttitle{Main Belt Yarkovsky}
\shortauthors{Hung et al.}

\begin{abstract}
We attempt to detect a signal of Yarkovsky-related acceleration in the orbits of 134 main belt asteroids (MBAs) we observed with the University of Hawai'i 88 inch telescope, supplemented with observations publicly available from the Minor Planet Center and Gaia Data Release 3. We estimated expected Yarkovsky acceleration values based on parameters derived through thermophysical modeling, but we were not able to find any reliable detections of Yarkovsky in our sample. Through tests with synthetic observations, however, we estimated the minimum observational arc length needed to detect the Yarkovsky effect for all of our sample MBAs, which in nearly every case exceeded the current arc length of the existing observations. We find that the Yarkovsky effect could be detectable within a couple of decades of discovery for a 100 m MBA assuming 0.$\arcsec$1 astrometric accuracy, which is at the size range detectable by the upcoming Vera Rubin Observatory Legacy Survey of Space and Time.
\end{abstract}

\section{Introduction}

The orbital elements of asteroids are constantly changing, subject to perturbations due to gravity, relativity, and radiation forces \citep[see the review by][]{Farnocchia15IV}. Asteroid orbits are chiefly defined by the gravitational force of the Sun, but the perturbations on that force must be accounted for in order to fit well-constrained orbits with all observational data points. It is standard practice to include the gravity of the planets, Moon, and Pluto, as well as large bodies in the asteroid belt, in the dynamical models for computing orbit solutions \citep[e.g.,][]{Farnocchia15IV}. Similarly, a general relativistic model for the Sun is sufficient to describe most asteroid orbits, but the relativistic contributions of the planets can be significant in the case of close encounters \citep[e.g.,][]{Chesley14}.

Although it imparts a much smaller acceleration than solar and planetary gravitational forces, one thermal force that can be important to consider is the Yarkovsky effect, which arises from an anisotropy in an asteroid's reemission of absorbed solar radiation \citep[see the reviews by][]{Bottke06, Vokrouhlicky15}. As asteroids are rotating bodies with nonzero thermal inertia, there is a delay between when an asteroid absorbs and reradiates heat from the Sun. Between these two events, the asteroid rotates by some angle, producing a net force along the asteroid's orbit. This transverse force induces a secular change on the asteroid's semimajor axis, causing the asteroid to drift outward or inward depending on the obliquity of its spin axis. The effect is maximized for obliquities of 0$\degree$ or 180$\degree$.

Generally, the Yarkovsky force is stronger for smaller asteroids close to the Sun due to the more favorable area-to-mass ratios and increased solar flux. However, if the diameter becomes so small that the thermal wave can penetrate the entire body, the day- and nightsides will equalize in temperature and weaken the Yarkovsky effect. While thermal inertia is necessary for the Yarkovsky force to manifest, very high values similarly dampen its effect. If the thermal inertia is so high that no heat is transferred over one rotation cycle, the temperature distribution remains uniform along lines of constant latitude; thus, there is no anisotropy in the thermal emission. The same is true for asteroids with zero or infinitely fast rotation \citep{Bottke06}.

Unlike gravity or relativity, nongravitational forces such as the Yarkovsky effect depend on the physical properties of the asteroid itself, which are generally unknown. The Yarkovsky effect can, however, be observed as an orbital deviation in the orbits of asteroids, but because of its small magnitude, it is only detectable in the case of very precisely known orbits, constrained by sufficiently long arcs of optical astrometry \citep{Vokrouhlicky15}. When feasible, radar observations greatly improve orbital constraints, as they provide information orthogonal to what we obtain with optical observations and at much higher precision \citep{Ostro02}, but they are limited to targets sufficiently close to Earth. 

Despite the currently hundreds of reported Yarkovsky detections among the near-Earth asteroid (NEA) population \citep[e.g.,][]{Chesley03, Chesley08, Chesley16, Vokrouhlicky08, Nugent12, Farnocchia13, Greenberg20}, the Yarkovsky effect has never been detected in main belt asteroids (MBAs) on an individual level through orbit determination. In particular, the increased heliocentric distance for MBAs relative to NEAs presents several key challenges. Due to the drop-off in apparent magnitude with heliocentric distance and size, it is more difficult to detect MBAs of the same size as NEAs. Thus, the MBAs that have been discovered are generally larger, so there are fewer viable candidates to investigate for signatures of the Yarkovsky effect compared with NEAs due to the inverse relationship between asteroid size and the Yarkovsky drift rate.

Signs of the Yarkovsky effect's influence in the main belt are, however, well-established, particularly as a mechanism for explaining the structure of asteroid families, which are the leftover fragments born from the catastrophic disruptions of parent asteroids. Members of the same family can be identified through commonalities in their proper orbital elements \citep{Milani94}. Asteroid families are observed to be sharply bounded by orbital resonances and may be asymmetric in proper element space, which can be explained by the semimajor axis drift induced by the Yarkovsky effect, shifting members further out from their parent asteroid until they fall into secular or mean-motion resonances that can boost their eccentricities and inclinations or eject them into planet-crossing orbits \citep{Bottke01}. The ages of asteroid families can be estimated by numerically integrating their members backward in time, which can be further refined by explicitly taking the Yarkovsky-induced orbital drift into account \citep[e.g.,][]{Nesvorny02, Nesvorny04}. Because the Yarkovsky effect is inversely proportional to asteroid size, the smallest members of a family end up furthest away from the parent body, displaying a characteristic ``V shape'' when plotting the absolute magnitude $H$ of the members against their semimajor axes \citep{Vokrouhlicky06a, Vokrouhlicky06b}. Family members drift either outward or inward depending on the sign of their rotation, as confirmed by the observed distributions \citep{Durech23}.

Searches for signatures of the Yarkovsky effect are often purely based on astrometric data, which allows one to ignore how the relevant physical properties are either poorly constrained or completely unknown for the vast majority of asteroids. It is possible to derive these properties by way of a thermophysical model \citep[TPM; e.g.,][]{Delbo15}, which requires both thermal infrared flux measurements and a shape model for the asteroid. However, shape models are only available for a few thousand asteroids, and many are crude approximations with unrealistic sharp edges \citep{Durech10}. Such poor shape models are physically unrealistic and likely to introduce inaccuracies into the thermophysical modeling \citep[e.g.,][]{Hanus15}. Even so, with some knowledge about the physical properties of a sample of asteroids in hand, it becomes possible to identify the most promising candidates for detecting Yarkovsky acceleration and therefore prioritize future observations to better constrain their orbits. Indeed, we already have one such successful case with the 0.5 km NEA (101955) Bennu, where its orbit was tightly constrained thanks to tracking data obtained by the sample return mission Origins, Spectral Interpretation, Resource Identification, Security, Regolith Explorer \citep[OSIRIS-REx;][]{Lauretta17}, which in turn validated the thermophysical modeling of its Yarkovsky acceleration \citep{Farnocchia21}.

\section{Yarkovsky Modeling}\label{sec.yarkmod}

Because the Yarkovsky effect primarily manifests as a semimajor axis drift, the nongravitational transverse acceleration is commonly described with a comet-like model \citep{Marsden73}, $a_t = A_2 g(r)$, where $g(r)$, originally formulated to describe \ce{H2O}-driven outgassing, is equal to $(1\,\mathrm{au}/r)^{2}$, and $A_2$ is a free parameter. As of 2023 September 21, there are published $A_2$ values for 340 NEAs in the Jet Propulsion Laboratory (JPL) Small-Body Database\footnote{\url{https://ssd.jpl.nasa.gov/tools/sbdb_query.html}}.

The complexity of Yarkovsky models can vary based on the amount of information available for a particular asteroid \citep[e.g.,][]{Vokrouhlicky00, Chesley14}. We use the fairly simple \citet{Farnocchia13} formulation for our work, which describes the $A_2$ parameter as

\begin{equation}\label{eq.yark}
A_2 = \frac{4(1-A)}{9} \Phi (1\,\mathrm{au}) f(\Theta) \cos (\gamma)
\end{equation}

\noindent where $A$ is the Bond albedo, and $\gamma$ is the spin obliquity, which was derived through lightcurve inversion along with the shape model. $\Phi (1\,\mathrm{au})$ is the standard radiation force factor, equal to

\begin{equation}
\Phi (1\,\mathrm{au}) = \frac{3 G_S}{2 \rho D c}
\end{equation}

\noindent where $G_S$ is equal to 1361 W m$^{-2}$ \citep{Kopp11} and is the solar constant at 1 au, $\rho$ is the bulk density, $D$ is the mean diameter, and $c$ is the speed of light. $f(\Theta)$ is a function of the thermal parameter $\Theta$, which is equal to

\begin{equation}
f(\Theta) = \frac{0.5 \Theta}{1 + \Theta + 0.5 \Theta^2}
\end{equation}

\noindent and $\Theta$ is given by

\begin{equation}
\Theta = \frac{\Gamma}{\epsilon \sigma_{SB} T_*^3} \sqrt{\frac{2 \pi}{P}}
\end{equation}

\noindent where $\Gamma$ is the thermal inertia, $\epsilon$ is the bolometric emissivity assumed to be 0.9, $\sigma_{SB}$ is the Stefan-Boltzmann constant, and $P$ is the rotational period. $T_*$ is the subsolar temperature, given by

\begin{equation}
T_* = \sqrt[4]{\frac{(1-A)G_S}{\epsilon \sigma_{SB} p^2}}
\end{equation}

\noindent where $p$ is the semilatus rectum, given by

\begin{equation}
p = a(1-e^2)
\end{equation}

\noindent where $e$ is the eccentricity, and $a$ is the semimajor axis.

\section{Data}
\subsection{Thermophysical Modeling Sample Selection}\label{sec.tpm}

\citet{Hung22} identified a total of 2551 asteroids observed by the Wide-field Infrared Survey Explorer \citep[WISE;][]{Wright10} that had both sufficient thermal data and existing shape models on the publicly available Database of Asteroid Models from Inversion Techniques \citep[DAMIT;][]{Durech10}. We then thermophysically modeled these asteroids and derived thermal inertia, diameters, and albedos for 1847 asteroids after making our $\chi^2$ quality cuts. We used these thermal parameters to compute an approximate Yarkovsky acceleration prediction for every asteroid we thermophysically modeled using the framework in \citet{Farnocchia13}, and we summarize this process below. We refer the reader to \citet{Hung22} for more details.

In our thermophysically modeled sample, only the NEAs (1685) Toro and (1865) Cerberus have explicitly determined $A_2$ values. Toro\footnote{\url{https://ssd.jpl.nasa.gov/tools/sbdb_lookup.html\#/?sstr=1685}} has $A_2 = -3.05 \pm 0.46 \times 10^{-15}$ au day$^{-2}$, and Cerberus\footnote{\url{https://ssd.jpl.nasa.gov/tools/sbdb_lookup.html\#/?sstr=1865}} has $A_2 = -10.22 \pm 3.93 \times 10^{-15}$ au day$^{-2}$. In order to identify the most promising Yarkovsky candidates for observational follow-up, we use our derived physical parameters to compute an approximate Yarkovsky acceleration prediction for every asteroid in our sample using Equation \ref{eq.yark}. We are not able to derive any information about the bulk density from thermophysical modeling, so we instead estimate the parameter using the average value expected from the asteroid's taxonomy \citep{Krasinsky02, Carry12}. In cases where the taxonomy was unknown, the taxonomy was assumed based on the asteroid's geometric albedo $p_V$. As a simplification, we assumed no uncertainties on the bulk densities we adopted, but our $A_2$ estimation would benefit from more careful consideration of the bulk density uncertainties should we find any valid Yarkovsky detections in our sample. We refer the reader to table 2 in \citet{Hung22} for the full list of values used.

The uncertainty on $A_2$ was determined by what values of $\Gamma$, $D$, and $A$ would maximize or minimize $A_2$ in the range of each parameter's respective TPM-derived 1$\sigma$ uncertainties. The rotation period and spin axis of each asteroid were determined through lightcurve inversion. We assumed no uncertainties on $\rho$, $P$, or $\gamma$. It is important to note that the thermal parameter uncertainties will be underestimated, as the TPM does not account for any uncertainties in the shape model and spin axis. In many cases, we could not constrain an asteroid's lower bound on thermal inertia in our TPM; thus, the lower bound on $A_2$ was similarly unconstrained.

\subsection{UH88 Observing Campaign}

\begin{figure}
\centering
\includegraphics[width=\columnwidth]{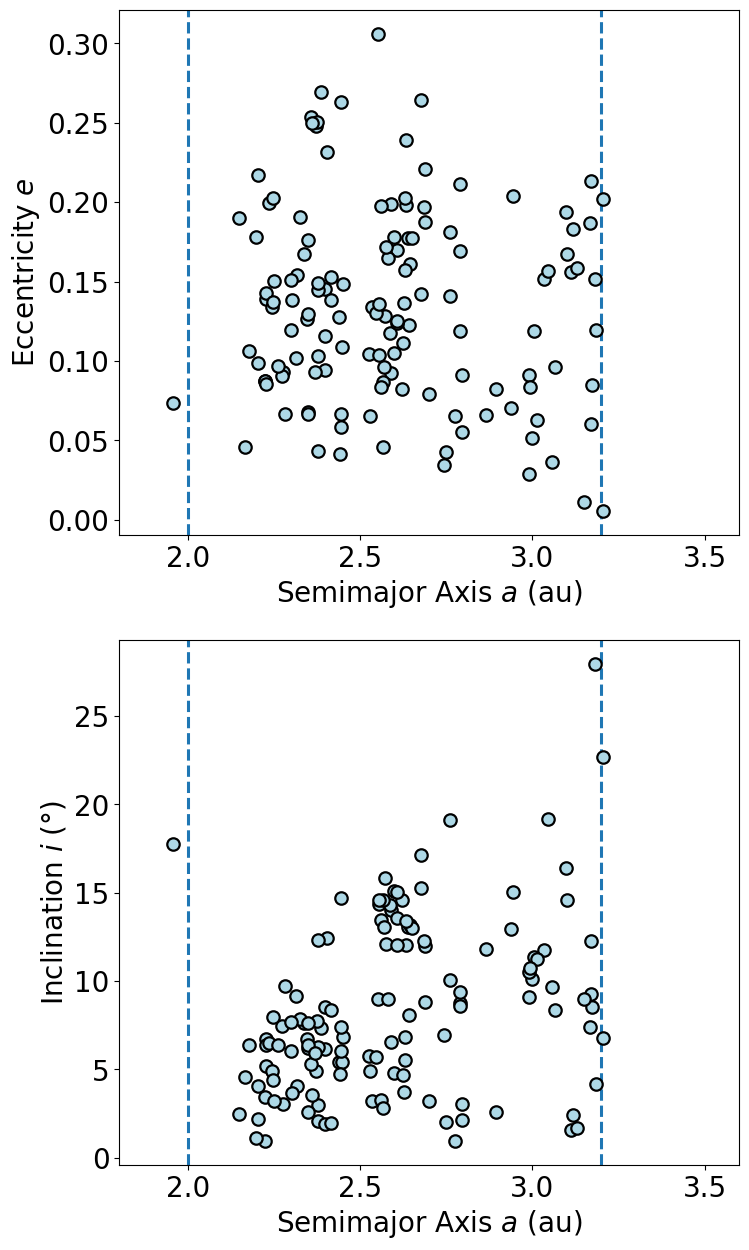} 
\caption{Semimajor axes, eccentricities, and inclinations of our 134 MBA targets. The vertical lines in semimajor axis are drawn at the boundaries of the inner ($a<2.0$ au), central ($2.0<a<3.2$ au), and outer ($3.2<a<4.6$ au) main belt regions. Nearly all of our observed MBAs fall within the central main belt region.}
\label{fig.eai}
\end{figure}

We obtained a total of 6 nights between 2021 April and 2022 February for visible-wavelength follow-up observations with the Semiconductor Technology Associates 10k charge-coupled device (CCD) camera (STACam) on the University of Hawai'i 88 inch (UH88) telescope, observing a total of 134 MBAs (Fig. \ref{fig.eai}), with 35 having diameters of at most 5 km, as well as, incidentally, one Mars-crosser (Table \ref{tab.obs}). We selected our targets out of our full thermophysically modeled sample based on what was observable on each night, with higher priority given to targets that had higher-quality thermophysical modeling fits and larger $A_2$ estimates. As a point of comparison, we also observed the two NEAs in our thermophysically modeled sample with a previously found Yarkovsky detection: (1685) Toro (JPL\footnote{\url{https://ssd.jpl.nasa.gov/tools/sbdb_lookup.html\#/?sstr=1685}}) and (1865) Cerberus (JPL\footnote{\url{https://ssd.jpl.nasa.gov/tools/sbdb_lookup.html\#/?sstr=1865}}).

Our targets were bright ($V \leq 20$ mag in the majority of cases) and generally observed below 1.5 airmass. While we used no filter for the first 3 nights of our observing run, we switched to using the $R$ band for the remaining nights in order to reduce the effects of differential color refraction on the astrometry, which become more significant the further away an observation is from zenith. At airmasses greater than 2, chromatic corrections may be as large as 0.$\arcsec$1 for white light \citep{Tholen18}. The use of any filter that narrows the bandpass will reduce the differential color refraction due to restricting the total wavelength range received by the camera. With the $R$-band in particular, we have the advantage of higher quantum efficiency and better seeing compared with shorter-wavelength filters. The filter also compensated for the somewhat nonuniform color response of the detector, which in turn simplified the flattening process.

We observed each target with a minimum of three dithered exposures to confirm its motion and identity. We set the exposure time to obtain a signal-to-noise ratio (S/N) of at least 10. As nearly all of our targets had apparent magnitudes of $V=20$ mag or brighter, this requirement was achievable with exposures much shorter than a minute for most asteroids. In some cases, however, we discarded exposures due to factors such as issues with telescope tracking inconsistencies, exceptionally poor seeing, or interference between the target and background stars. Additionally, some of our targets were located near the center of the galactic plane, which is home to very dense star fields and thus many potential places for the flux of a background star to abut or completely surround an asteroid. Although we attempted to take observations of this region of the sky, we discarded the observations for a small number of asteroids where we were unable to locate the target within the field.

Ideally, we would have a minimum of two internally consistent observations on different nights to confirm any detection of Yarkovsky drift, but weather conditions precluded second-night observations on several targets. We nonsidereally tracked each target, matching the motion of the target. However, our MBA targets moved slowly enough to keep the background stars as approximate point sources rather than trails in the exposure times we used. The NEAs moved about an order of magnitude faster and thus required a trailed model for the stars, though this had a negligible effect on the accuracy of the astrometry.

\begin{deluxetable*}{rlccccccr}
\tablecolumns{9}
\tablecaption{UH88 Observations\label{tab.obs}}
\tablehead{\twocolhead{Designation} & \colhead{2021} & \colhead{2021} & \colhead{2021} & \colhead{2021} & \colhead{2021} & \colhead{2022} & \colhead{Predicted $A_2$} \\
           \twocolhead{} & \colhead{Apr 13} & \colhead{May 12} & \colhead{Jun 10} & \colhead{Sep 25} & \colhead{Oct 23} & \colhead{Feb 27} & \colhead{($10^{15}$ au day$^{-2}$)} }
\startdata
767 & Bondia & \nodata & \nodata & \nodata & \nodata & \nodata & \checkmark & $0.07^{+0.01}_{-0.02}$ \\ 
769 & Tatjana & \nodata & \nodata & \nodata & \nodata & \nodata & \checkmark & $0.16^{+0.03}_{-0.16}$ \\ 
963 & Iduberga & \checkmark & \checkmark & \checkmark & \nodata & \nodata & \nodata & $0.93^{+0.10}_{-0.26}$ \\ 
1169 & Alwine & \checkmark & \checkmark & \checkmark & \nodata & \nodata & \nodata & $-0.91^{+0.64}_{-0.54}$ \\ 
1415 & Malautra & \nodata & \checkmark & \checkmark & \nodata & \nodata & \nodata & $0.25^{+0.10}_{-0.16}$ \\ 
\vspace{-0.3cm}
\enddata
\tablecomments{Observations for each target asteroid separated by date. We observed a total of 134 unique MBAs, as well as one Mars-crosser (2204 Lyyli) and two NEAs (1685 Toro and 1865 Cerberus). Each night consists of at least three dithered exposures per target. Due to the use of multiple shape models in the thermophysical modeling, there is often more than one set of derived thermal properties for an asteroid. The predicted $A_2$ reported here uses the thermal inertia, diameter, and albedo derived from the thermophysical modeling fit with the smallest $\chi^2$ for that asteroid. Refer to \citet{Hung22} for more details. The full version of this table can be found in the appendix in Table \ref{tab.obs2}.}
\end{deluxetable*}
\vspace{-9mm}

We overscan-subtracted, bias-corrected, and flat-fielded the STACam data according to the standard reduction process. We performed the bias correction using a master bias frame generated from a median combination of 15 bias frames taken at the start of each night. We generated the flat field from a median combination of all science images, excluding the images with especially dense star fields. A custom flat field is necessary for targets near the Moon, as the moonlight will create a gradient in the sky background. For the 2 nights where significant moonlight was a concern, we took an average of 12 dithered exposures for any target with a lunar elongation of less than 50$\degree$, which then we median combined to form the custom flat field. 

We visually identified the position of the target asteroid in each image, and we matched the reference stars against the Gaia Data Release 2 \citep[DR2;][]{Gaia18} catalog using the astrometric software \texttt{AstroMagic}\footnote{\url{http://www.astromagic.it/eng/astromagic.html}}. The matches made over STACam's 14.5 arcmin$^2$ field of view ranged between around 100 and over 10,000 in the cases of the dense star fields.

We then determined the final positions of the reference stars and target asteroid through a centroiding process, first run using a large aperture to accommodate errors in the \texttt{AstroMagic} pattern matching (e.g., distortions caused by refection) and then again using the optimum aperture, automatically determined by what matched the seeing. In each pass, we fitted a Gaussian distribution to the point-source image profile for each matched reference star and the target asteroid after a background subtraction, where we took the background to be the median value of the pixels in the annulus surrounding the aperture. The least-squares fit was repeated until convergence, which usually happened in about four iterations. Finally, the astrometric fitting was performed over six passes in order to eliminate astrometric and photometric outliers (brought about by, e.g., multiple sources in the same aperture), though all astrometric outliers were usually identified in a single pass. We performed the astrometric fits linearly, as the size of the detector was small enough to ignore higher-order contributions.

The centroiding of the target asteroid was occasionally corrupted due to other sources that were misattributed to the asteroid's flux. We corrected single bad pixels brought about by cosmic-ray hits or instrument defects by masking. We removed the flux from an abutting background star by a self-subtraction in the image, which involved making a copy of the image. We selected a nearby and slightly brighter isolated star and scaled the flux down in the copied image to match the abutting star. We aligned the abutting star in the original image with the scaled isolated star in the copied image and performed the subtraction. We then performed the centroiding again on the newly isolated asteroid. 

We submitted our astrometry to the Minor Planet Center (MPC)\footnote{\url{https://minorplanetcenter.net/}}, with average target asteroid uncertainties on R.A. and decl. positions of 0.$\arcsec$03 and ranging between 0.$\arcsec$01--0.$\arcsec$36. The standard deviation was 0.$\arcsec$03 in R.A. and 0.$\arcsec$02 in decl. Unmodeled sources of error include variations in seeing, transparency, and tracking within the same exposure, but these issues become less of a concern when there is minimal trailing in the background stars, such as in our observations. 

\section{Orbit Determination}

For well-constrained orbits, the orbital deviations caused by the Yarkovsky effect become visible in the data. We computed our orbit solutions with the JPL asteroid and comet orbit determination software. Its methods are described in \citet{Farnocchia15IV}. An assumed set of orbital elements and dynamical model will define an initial orbit that is fit to a set of observations. The best fit is determined by the orbit that minimizes the sum of the squares of the residuals, i.e., the differences between the measured and computed sky position for each observation. 

In addition to our own observations with UH88, we also used observations from Gaia Data Release 3 \citep[DR3;][]{Tanga22}, which includes astrometry for over 150,000 solar system objects among its extensive body of data products collected between 2014 July 25 and 2017 May 28. Gaia rotates at a constant rate \citep{Gaia16}, so the transit of a source on the focal plane will end up crossing nine CCD strips. A single transit may thus have up to nine total positions, one for each CCD. The astrometric positions have both per-transit systematic errors and individual random errors. We selected the first position per transit, and the corresponding correlated uncertainty was obtained by adding the systematic and random covariances.

The remaining observations we used come from those reported to the MPC, retrieved on 2023 June 1. The MPC serves as the central database for positional measurements of asteroids and other small bodies. Observation records for MBAs can span time frames over a century. However, most observations obtained from the MPC were reported using a format that did not allow for any communication of astrometric uncertainties. Thus, some assumptions must be made on the uncertainties of each reported observation when computing an orbit, which is handled by way of a weighting scheme. The weights are determined based on statistical analyses of the astrometric errors of past observations, which may be separated by variables such as the historical accuracy of the observing site, reported magnitude, and epoch of observation \citep[e.g.,][]{Chesley10, Farnocchia15, Veres17}. We used the debiasing scheme of \citet{Eggl20} and the weighting scheme of \citet{Veres17} for our orbit determination computations, which are, respectively, the most recent debiasing and weighting schemes available. Outliers were identified and rejected with the scheme described in \citet{Carpino03}. 

\subsection{Force Model}

The gravitational accelerations of the Sun, eight major planets, Pluto, and Moon are accounted for using the JPL planetary and lunar Development Ephemeris DE441 \citep{Park21}. We also include the gravitational contributions of the 16 most massive perturbers in the main asteroid belt described in the small-body perturber file SB441-N16 \citep{perturbers}.

The relativistic contributions of the Sun, planets, and Moon are included through the Einstein--Infeld--Hoffmann equations of motion, which approximate the dynamics of a system of pointlike masses due to mutual gravitational interactions, as well as general relativistic effects. The equations of motion are described in a first-order post-Newtonian expansion as detailed in \citet{Einstein38}, \citet{Will93}, and \citet{Moyer03}.

The Yarkovsky model is the \citet{Farnocchia13} formulation, which is what is described in \S\ref{sec.yarkmod}.

\subsection{Determining Valid Detections of $A_2$}

Plausible detections of $A_2$ are defined by the S/N and the indicator parameter $\mathcal{S}$, which is the ratio between the $A_2$ derived by the orbit determination and the expected $A_2$, typically taken from diameter scaling the $A_2$ of a reference asteroid with a reliable Yarkovsky detection \citep[e.g.,][]{Farnocchia13, Vokrouhlicky15, DelVigna18}. Because we have some knowledge of our asteroid sample's physical properties thanks to our thermophysical modeling efforts, we can instead use the per-asteroid predicted $A_2$ values as described in \S\ref{sec.tpm}. 

We consider detection to be valid if it has an S/N of at least 3 with a $\mathcal{S}$ that is close to 1. Cases with $\mathcal{S} \gg 1$ could indicate instances of nongravitational acceleration too strong to be explained by the Yarkovsky effect or, more likely, be spurious detections as a result of a poor orbital fit. Such cases could also imply a much lower bulk density or smaller size for the asteroid than what was assumed. Cases with $\mathcal{S} \ll 1$ would similarly suggest that our predicted $A_2$ values are overestimated as a result of inaccuracies in the values used for one or more of the asteroid's physical parameters, where the Yarkovsky force weakens with larger sizes, higher densities, very high or very low thermal inertia, and obliquities closer to 90$\degree$. However, low-$\mathcal{S}$ cases can still be valid detections of Yarkovsky acceleration and could offer new constraints on the physical properties of the asteroid \citep{Vokrouhlicky15}. We adopt the filtering criteria of \citet{DelVigna18} in only accepting detections with both S/N $> 3$ and $\mathcal{S} \leq 2$ as valid.

\section{Results}\label{sec.nodetect}

With the existing observations, we were unable to find any plausible detections of Yarkovsky acceleration in our observed MBA sample, suggesting that the astrometry was not of sufficient accuracy or observational arc length (Table \ref{tab.a2s}). Every MBA orbit solution produced an $A_2$ value at very low S/N, and many values were much larger than predicted. 

\begin{figure}
\centering
\includegraphics[width=\columnwidth]{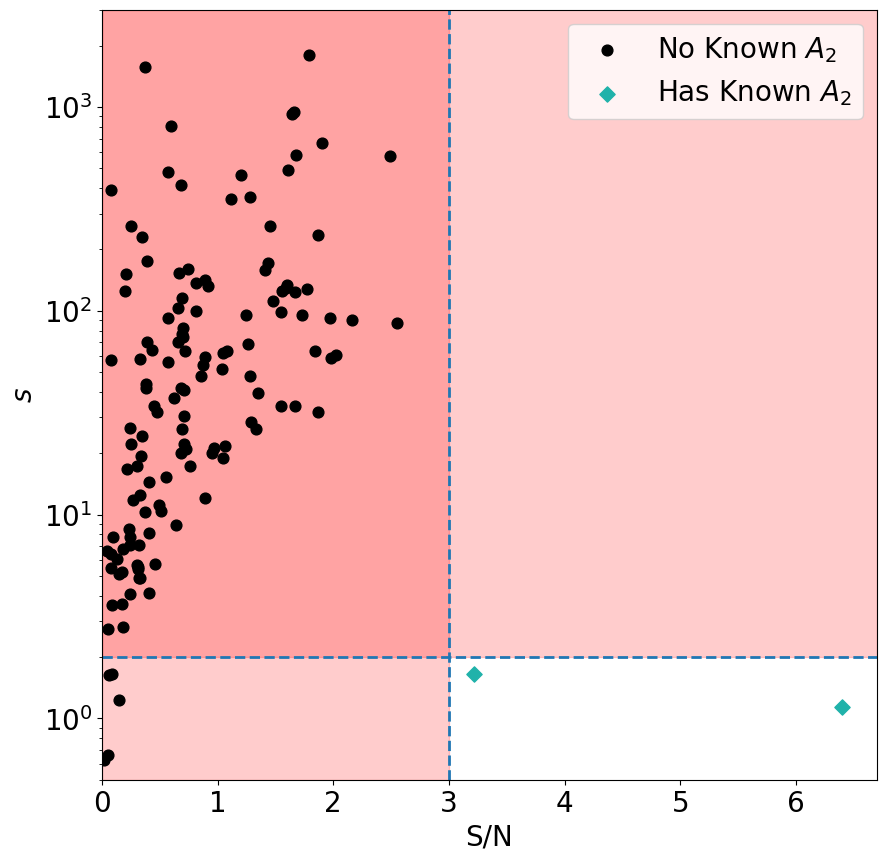}
\caption{$A_2$ derived by orbit determination for our sample of 134 MBAs, one Mars-crosser, and two NEAs plotted in terms of the S/N and indicator parameter $\mathcal{S}$. We define the $A_2$ to be valid if it has both S/N $> 3$ and $\mathcal{S} \leq 2$. All of our MBAs fall outside of this range, while the two NEAs are our only valid detections. Notably, the two NEAs are also the only asteroids in our sample with a previously found $A_2$ in the literature.}
\label{fig.snr_s}
\end{figure}

We did, however, find valid detections with the two NEAs in our sample that already had previously found $A_2$ detections in the literature (Fig. \ref{fig.snr_s}). The NEAs, (1685) Toro and (1865) Cerberus, show good agreement with the existing estimates. Toro is reported to have an $A_2$ of $-3.05 \pm 0.46 \times 10^{-15}$ au day$^{-2}$ in the JPL Small-Body Database\footnote{\url{https://ssd.jpl.nasa.gov/tools/sbdb_lookup.html\#/?sstr=1685}}. We find an $A_2$ of $-2.95 \pm 0.46 \times 10^{-15}$ au day$^{-2}$, which translates to an S/N of 6.4 and is within 1$\sigma$ of the JPL value. Cerberus is reported to have an $A_2$ of $-10.22 \pm 3.93 \times 10^{-15}$ au day$^{-2}$ on the JPL Small-Body Database\footnote{\url{https://ssd.jpl.nasa.gov/tools/sbdb_lookup.html\#/?sstr=1865}}. Our orbit determination gives us an $A_2$ of $-11.90 \pm 3.69 \times 10^{-15}$ au day$^{-2}$, corresponding to an S/N of 3.2, which is again within a 1$\sigma$ agreement. The $A_2$ value determined can be rather sensitive to the astrometric data set used, coming down to the choice of what observations are included and at what weights, but the close agreement suggests that the differences are minor.

Toro is the only asteroid in our sample with the benefit of existing radar observations, which is partially why its $A_2$ detection is of higher S/N than that of Cerberus, which has no radar observations and a slightly shorter observational arc. Cerberus was originally found with a somewhat weak Yarkovsky acceleration signal of 2.1$\sigma$ by \citet{Greenberg20} using observations between 1971 and 2019. Including our UH88 observations and Gaia DR3, as well as observations from a host of other sites, Cerberus's observational arc is now 2 yr longer. As a consequence, we have found an improvement to the significance of the Yarkovsky acceleration detection of 1.1$\sigma$, illustrating the importance of having accurate observations for such searches.

\begin{deluxetable*}{rlrccc}
\tablecolumns{7}
\tablecaption{$A_2$ Values from Orbit Determination\label{tab.a2s}}
\tablehead{\twocolhead{Designation} &  \colhead{Predicted $A_2$} & \colhead{Derived $A_2$} & \colhead{S/N} & \colhead{$\mathcal{S}$} \\
           \twocolhead{} & \colhead{($10^{15}$ au day$^{-2}$)} & \colhead{($10^{15}$ au day$^{-2}$)} & \colhead{} & \colhead{}}
\startdata
   1685 & Toro             & $-2.60^{+0.31}_{-0.19}$ & $  -2.95 \pm   0.46$ & 6.40 &    1.1 \\ 
   1865 & Cerberus         & $-7.18^{+0.85}_{-0.57}$ & $ -11.90 \pm   3.69$ & 3.22 &    1.7 \\ 
   1773 & Rumpelstilz      & $ 0.77^{+0.05}_{-0.09}$ & $ -67.23 \pm  26.41$ & 2.55 &   87.2 \\ 
  22092 & 2000\,AQ199      & $-0.00^{+0.00}_{-0.23}$ & $-132.11 \pm  52.99$ & 2.49 &  575.1 \\ 
   1430 & Somalia          & $ 0.70^{+0.16}_{-0.70}$ & $ -63.42 \pm  29.32$ & 2.16 &   90.3 \\ 
\vspace{-0.3cm}
\enddata
\tablecomments{Solutions for our 134 MBAs, one Mars-crosser (2204 Lyyli), and two NEAs (1685 Toro and 1865 Cerberus). For a detection to be valid, we require both S/N $> 3$ and $\mathcal{S} \leq 2$, where $\mathcal{S}$ is the absolute value of the ratio of the derived and predicted $A_2$. In cases where the predicted $A_2$ was nominally zero, we used the larger error bar value for this calculation instead. Only the NEAs can be considered valid Yarkovsky detections. The remaining MBAs had $A_2$ values associated with very low S/N which were often orders of magnitude larger than expected. The full version of this table can be found in the appendix in Table \ref{tab.a2s2}.}
\end{deluxetable*}
\vspace{-9mm}

While our two NEA $A_2$ detections fall within 1$\sigma$ agreement with the literature comparisons, we note that this is a much too limited sample size from which to draw any major conclusions. The slight differences between our determined $A_2$ values and those in the literature come about mainly due to the differences in the choice of what observations were included in the orbit solution and how the data were weighted. The earliest observations in particular can be highly influential in the orbit solution, as they define the total length of the observational arc, but they are often isolated in time, sometimes separated by several years or decades from the next earliest observations. The accuracy and precision of the earliest observations are also much more suspect compared to modern observations thanks to general technological improvements over time in instruments, cameras, and star catalogs, particularly for observations taken before 1950 \citep{Veres17}.

Although weighting schemes can statistically account for the expected accuracy of an observation based on factors such as the observing site or epoch, old data are sparse, which precludes any reliable statistical schemes to correct whatever biases they might have. The star catalog is also often unknown for old observations, which can lead to additional arcseconds of error, as the data cannot be debiased. Ideally, it would be best to manually reweight all suspect observations, but this can be a very time-consuming process with little benefit. It is much more important when working with a plausible new Yarkovsky detection to ensure that an extraordinary result is indeed real, but we have no such cases in our MBA sample. 

It is possible to improve the accuracy and precision of archival data by remeasuring against modern star catalogs, as was done for four 1953 precovery observations to confirm a Yarkovsky detection in the orbit of (152563) 1992\,BF \citep{Vokrouhlicky08}, but this may not always be feasible. These observations predate the adoption of the CCD in the late 20th century, which replaced photographic plates with a data format that could be electronically stored and processed \citep{Tokunaga14}. Observations from photographic plates could only be remeasured if the plates were first digitized, which can be a costly procedure and is dependent on the plates having been preserved in some form over the past several decades. Moreover, nothing can be done about the temporal gaps in the astrometry without coming across more archival data by happenstance, which will be less likely the older the observations are required to be.

Given the low quality of the early observational records, as well as the several variables the Yarkovsky effect depends on, it is difficult to constrain the parameter space with an unsuccessful detection beyond drawing the broadest of conclusions. Let us use the smallest $A_2$ uncertainty found among our MBAs as an example. We found an $A_2$ uncertainty of $14 \times 10^{-15}$ au day$^{-2}$ for (1518) Rovaniemi, which means a valid $A_2$ detection would be at minimum equal to $42 \times 10^{-15}$ au day$^{-2}$ in magnitude for an S/N of 3. For this exercise, we will consider our diameter of 8.4 km as robustly determined thanks to the agreement found with the TPM-derived values in \citet{Hung22} and adopt values for the Bond albedo, thermal inertia, and obliquity in Equation \ref{eq.yark} such that we maximize the Yarkovsky acceleration. $A_2$ is thus left to vary inversely with the bulk density, where we find that $A_2$ only reaches $42 \times 10^{-15}$ au day$^{-2}$ for a bulk density of 0.080 g cm$^{-3}$, too low to be physically plausible and a factor of 34 times smaller than the average S-type bulk density of 2.71 g cm$^{-3}$ we adopted for Rovaniemi. Further observations will be necessary for more useful constraints on the parameter space. 

\section{Discussion}
\subsection{Finding the Minimum Arc Length for Yarkovsky Detectability with Synthetic Observations}

The currently available observational data set precluded us from finding any concrete signal of the Yarkovsky effect in the MBAs in our sample, and the lack of detection in turn offered only very poor constraints on the parameter space of the relevant physical properties. We can, however, investigate what observations might be necessary to obtain a Yarkovsky detection using data simulated with the \texttt{OrbFit} software package\footnote{Version 5.0.7, \url{http://adams.dm.unipi.it/orbfit/}}. 

The parameter space for testing synthetic observations is enormous. We could consider the arc length, cadence, precision, and location on Earth of the observations. The orbital properties of the asteroid, such as its semimajor axis and eccentricity, also play an important role, as the difficulty of detecting $A_2$ increases with heliocentric distance. In order to simplify our investigation, we consider only the best-case scenario with current technology using a fixed observing strategy with a select few asteroids in order to find a quantified relation between arc length, semimajor axis, and the detectability of $A_2$.

We draw three MBAs from our observed sample to serve as our test cases. Restricting ourselves to low-eccentricity ($e<0.1$) asteroids, we choose the minimum, median, and maximum semimajor axes among them: (45898) 2000\,XQ49 at $a = 1.955$ au, (50219) 2000\,AL237 at $a = 2.567$ au, and (13936) 1989\,HC at $a = 3.201$ au. We use a range of semimajor axes to test the effects of differing angular size, and we use low-eccentricity asteroids to ensure the heliocentric distance is as consistent as possible. We vary the observational arc lengths starting at 1980 January 1 from 10 to 200 yr in steps of 10 yr with a fixed cadence of 50 days between the observations. For each observation, the observer is assumed to be geocentric. The uncertainty on R.A. and decl. is assumed to be 0.$\arcsec$1, a slightly more conservative value than the average uncertainties we found with our UH88 observations.

For each set of simulated observations, we used \texttt{OrbFit} to estimate $A_2$ and its corresponding uncertainty. The detectability of $A_2$ would simply be determined by the S/N, which is the $A_2$ value divided by the \texttt{OrbFit} $A_2$ uncertainty. Therefore, we can simply express the $A_2$ uncertainty as a function of arc length and semimajor axis. The S/N is then obtained by linearly scaling the $A_2$ uncertainty to whatever target $A_2$ value we choose. 

\subsection{Detectability Thresholds in Our Observed Sample}\label{sec.minarc}

\begin{figure*}
\centering
\includegraphics[width=2\columnwidth]{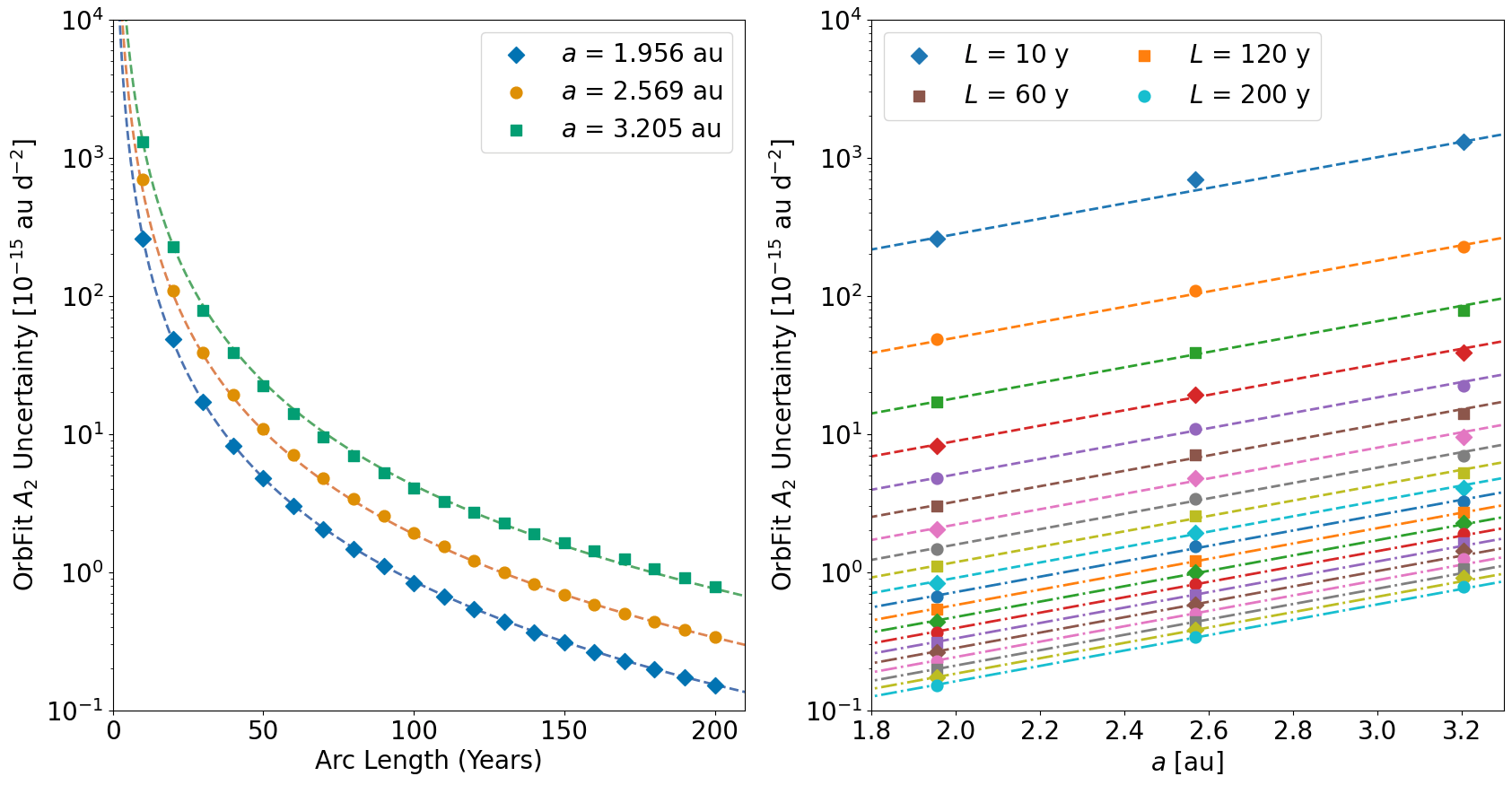} 
\caption{The $A_2$ uncertainty as a function of observational arc length $L$ and asteroid semimajor axis $a$. With the semimajor axis fixed in the left panel, the arc length $L$ and $A_2$ uncertainty $A_{2, unc}$ follow a power law. With the arc length fixed in the right panel, $a$ and $A_{2, unc}$ follow an exponential function. Note that the $A_2$ uncertainty decreases as the arc length increases from 10 to 200 yr in 10 yr steps, but we only include four arc lengths in the legend for the purpose of brevity. The fitted lines are plotted from our two-dimensional fit, $f(L, a) = k 10^{c (a-2.6)} L^\alpha$, where the values of the coefficients are $k=1.852 \times 10^5$, $c=0.556$, and $\alpha=-2.488$. }
\label{fig.unc_fit}
\end{figure*}

If we hold the semimajor axis constant, the relation between arc length $L$ and $A_2$ uncertainty $A_{2, unc}$ essentially follows a power law (Fig. \ref{fig.unc_fit}, left panel). If we hold the arc length constant, the semimajor axis $a$ and $A_{2, unc}$ appear to follow some sort of exponential function (Fig. \ref{fig.unc_fit}, right panel). We can thus combine the two and fit $A_{2, unc}$ with the two-dimensional function

\begin{equation}\label{eq.2dfit}
f(L, a) = k 10^{c (a-2.6)} L^\alpha
\end{equation}

\noindent where $k$, $c$, and $\alpha$ are our fitting coefficients. We use the \texttt{minimize} routine with the Nelder--Mead method \citep{neldermead} in the \texttt{Python} library \texttt{SciPy} to minimize the sum of the residuals $r$, which are given by

\begin{equation}
\sum_{i} r_i^2 = \sum_{i} \left (1 - f(L_i, a_i)/A_{2, unc_i} \right)^2
\end{equation}

We minimize the relative error on $A_{2, unc}$ rather than the absolute in order to avoid biasing the fit toward larger values of the $A_2$ uncertainty, which would consequently lead us to better fit shorter observational arcs than longer arcs. We find best-fit values of $k=1.852 \times 10^5$, $c=0.556$, and $\alpha=-2.488$. We stress for the reader's consideration that our fit here is only a proxy for the detectability of the Yarkovsky effect in the main belt.

\begin{figure}
\centering
\includegraphics[width=\columnwidth]{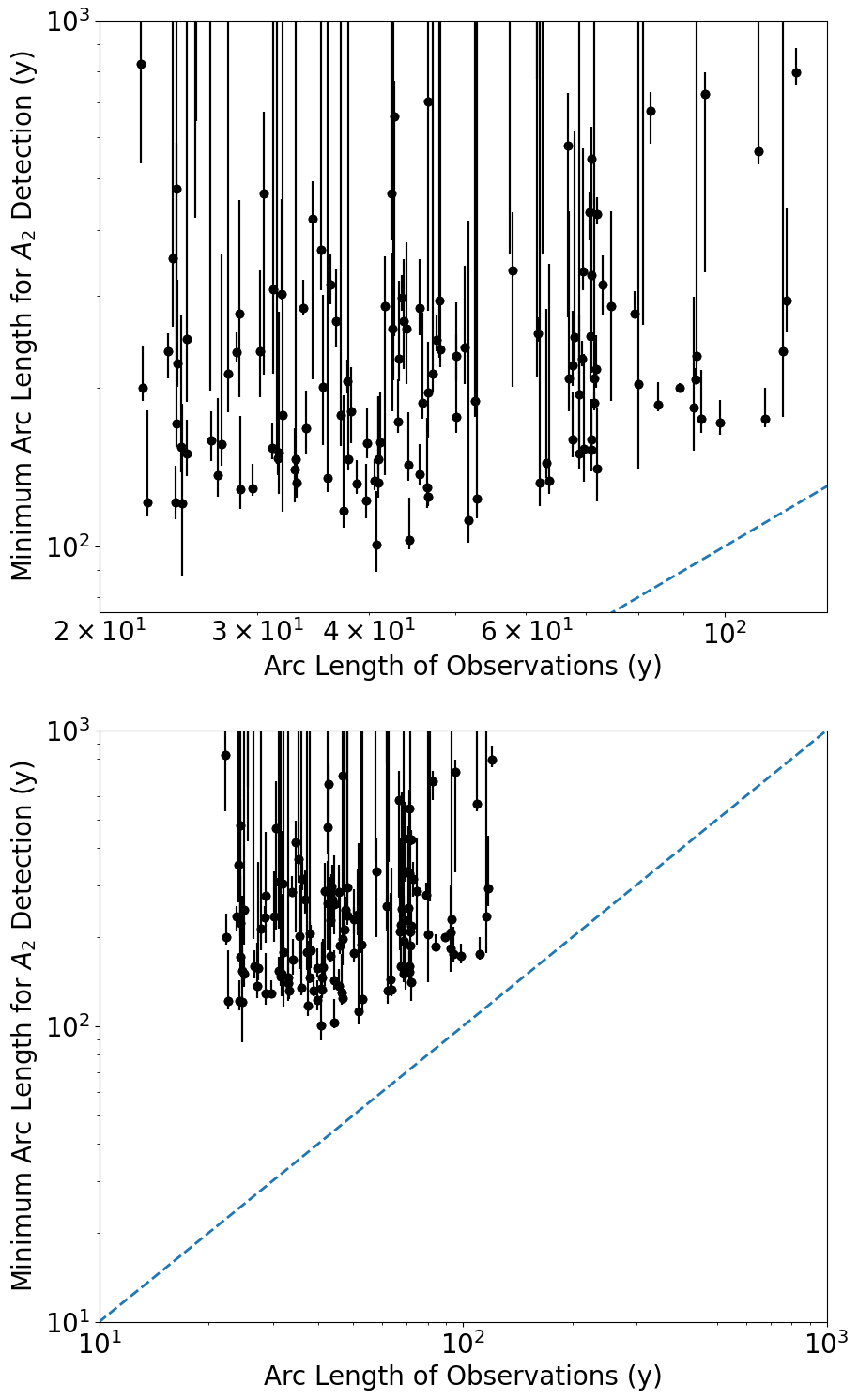}
\caption{We compare the observational arc lengths of our observed sample of MBAs with the minimum arc lengths needed for a Yarkovsky detection. The dashed line shows where the two axes are equal. The error bars on the minimum arc lengths are based on the error bars on the predicted $A_2$ values derived from our TPM fitting as described in \S\ref{sec.tpm}. They are best taken strictly as lower limits due to the caveats described in \S\ref{sec.minarc}. None of our observed asteroids have long enough arcs to meet the threshold for Yarkovsky detection. }
\label{fig.arc_compare}
\end{figure}

The semimajor axis is a known quantity for every asteroid in our observed sample. If we define the threshold for detectability to be S/N = 3, then the $A_2$ uncertainty must be a third of the target $A_2$ value. With this in mind, we can thus translate the predicted $A_2$ values of our sample into minimum arc lengths for detectability with Equation \ref{eq.2dfit} (Table \ref{tab.minarc}). If we compare our minimum arc lengths for Yarkovsky detectability to the arc lengths of the existing observations for our observed sample of MBAs, we see that no asteroids come close to meeting the threshold (Fig. \ref{fig.arc_compare}). Even if any had met the threshold, it is unlikely that we would have been able to find a successful Yarkovsky detection. The synthetic observations were modeled after the best-case scenario under modern-day standards, with a regular cadence and much higher levels of precision than were accessible a century ago \citep{Veres17}. Observations of lower precision will necessitate longer arc lengths for Yarkovsky detection to be possible for a given $A_2$ value. While (767) Bondia, for example, has the longest observational arc in our sample, with its earliest observations in 1902, there are multiple years-long gaps in its pre-1956 record, and many observations are only to arcminute precision\footnote{\url{https://minorplanetcenter.net/db_search/show_object?utf8=\%E2\%9C\%93&object_id=767}}. Finally, it is also important to note that this discussion is in part predicated on the accuracy of our predicted $A_2$ values, which can vary between asteroids due to differences in data quality. The predicted $A_2$ values were estimated from parameters derived through thermophysical modeling (see \S\ref{sec.tpm}), which becomes less reliable when applied to cruder shape models \citep{Hung22}. 

\begin{deluxetable*}{rlccrcc}
\tablecolumns{7}
\tablecaption{Minimum Arc Lengths Needed for Yarkovsky Detection\label{tab.minarc}}
\tablehead{\twocolhead{Designation} & \colhead{$a$} & \colhead{$e$} & \colhead{Predicted $A_2$} & \colhead{Min. $L$} & \colhead{Current $L$} \\
           \twocolhead{} & \colhead{(au)} & \colhead{} & \colhead{($10^{15}$ au day$^{-2}$)} & \colhead{(yr)} & \colhead{(yr)}}
\startdata
  767 & Bondia & 3.119 & 0.1829 & $0.07^{+0.01}_{-0.02}$ & 751--888 & 120 \\ 
  769 & Tatjana & 3.166 & 0.1870 & $0.16^{+0.03}_{-0.16}$ & 532\texttt{+} & 109 \\ 
  963 & Iduberga & 2.248 & 0.1378 & $0.93^{+0.10}_{-0.26}$ & 168--200 & 111 \\ 
 1169 & Alwine & 2.319 & 0.1549 & $-0.91^{+0.64}_{-0.54}$ & 152--299 & 92 \\ 
 1415 & Malautra & 2.223 & 0.0874 & $0.25^{+0.10}_{-0.16}$ & 255--441 & 117 \\ 
\vspace{-0.3cm}
\enddata
\tablecomments{Minimum observational arc lengths $L$ needed to detect a predicted $A_2$ value at S/N = 3 for our sample of 134 observed MBAs and one Mars-crosser (2204 Lyyli), calculated as described in \S\ref{sec.minarc}. The semimajor axis $a$ and eccentricity $e$ are taken from the JPL Small-Body Database. The predicted $A_2$ values are estimated from TPM-derived parameters as described in \S\ref{sec.tpm}. The current arc lengths are the total time span of the existing observations for each asteroid, last retrieved on 2023 January 1. The minimum arc length spans cover the $1\sigma$ range in the predicted $A_2$ values. In instances where the predicted $A_2$ reaches zero, the minimum arc length is essentially infinite; thus, we only report a lower bound for such asteroids. The full version of this table can be found in the appendix in Table \ref{tab.minarc2}.}
\end{deluxetable*}
\vspace{-9mm}

\subsection{Detectability Thresholds for Undiscovered MBAs}

The difficulty in detecting the Yarkovsky effect in the main belt is in part due to the sizes of known MBAs. An MBA of the same size as an NEA would be more difficult to detect due to the drop-off in apparent magnitude with heliocentric distance; thus, many fewer subkilometer MBAs are known relative to NEAs as a result of observational bias. However, we can expect asteroid discovery capabilities to dramatically improve in the near future with upcoming surveys such as the Vera Rubin Observatory Legacy Survey of Space and Time \citep[LSST;][]{lsst}, expected in 2024, which will be able to probe fainter magnitudes. LSST will have an 8.4 m primary mirror with a 9.6 deg$^2$ field of view and is expected to be capable of detecting asteroids with diameters in the range of 100 m at main belt distances. 

Previous Yarkovsky studies have used a relation scaling with the most reliable Yarkovsky detection to obtain an expected value for $A_2$ in cases where the physical characterization of an asteroid is unknown \citep[e.g.,][]{Farnocchia13, DelVigna18}. The Yarkovsky effect is inversely proportional to the asteroid diameter, so the scaling relation is of the form

\begin{equation}\label{eq.scale}
A_{2, \mathrm{exp}} = A_{2, \mathrm{ref}} \frac{D_{\mathrm{ref}}}{D}
\end{equation}

\noindent where $A_{2, \mathrm{exp}}$ is the expected $A_2$ for an asteroid of diameter $D$ scaled with the corresponding values of a reference asteroid. In order to keep the parameter space as simple as possible, we assume all other parameters that affect the Yarkovsky acceleration, such as the bulk density, obliquity, and albedo, to be the same as the reference asteroid. Currently, the most reliable detection is that of (101955) Bennu. Bennu was visited by the sample return mission OSIRIS-REx \citep{Lauretta17} from late 2018 to 2020, where the addition of the spacecraft's tracking data and physical characterization helped refine Bennu's Yarkovsky detection down to 0.07\% precision \citep{Farnocchia21}.

Bennu has an estimated semimajor axis drift rate of $\left \langle da/dt \right \rangle = -284.6 \pm 0.2$ m y$^{-1}$ \citep{Farnocchia21} and an equivalent diameter of 0.482 km \citep{Daly20}. Using equation 5 in \citet{Farnocchia13} with $a_{\mathrm{Bennu}} = 1.126$ au\footnote{\url{https://ssd.jpl.nasa.gov/tools/sbdb_lookup.html\#/?sstr=101955}}, $e_{\mathrm{Bennu}} = 0.204$, and $n_{\mathrm{Bennu}} = 0.824$ deg d$^{-1}$ and assuming $d=2$, this translates to an $A_{2, \mathrm{Bennu}}$ of $-45.57 \pm 0.03 \times 10^{-15}$ au day$^{-2}$. 

\begin{figure}
\centering
\includegraphics[width=\columnwidth]{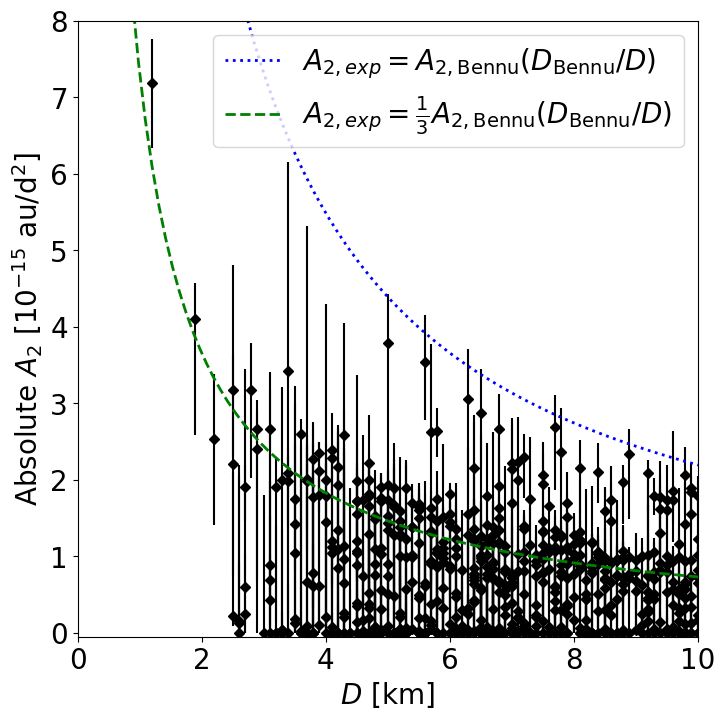}
\caption{The diameters of the nearly 2000 MBAs thermophysically modeled in \citet{Hung22} plotted against the absolute values of their predicted $A_2$, which was estimated from their derived parameters. The scaling relation from Equation \ref{eq.scale} using the values of Bennu produces consistently higher expected $A_2$ values than predicted, but we see better agreement if we reduce the relation by a factor of 3. }
\label{fig.a2_diam}
\end{figure}

It is important to note, however, that Bennu as an NEA may not necessarily be reflective of the general population of MBAs. Even among NEAs, Bennu is notable for its extreme obliquity and low bulk density \citep[$\sim$1 g cm$^{-3}$;][]{Farnocchia21}, which both maximize the Yarkovsky acceleration. Using Bennu as our reference asteroid consistently yields expected $A_2$ values that are higher than the $A_2$ values we estimated with our TPM-derived parameters. We can, however, find better agreement if we simply apply a factor of one-third to the scaling relation (Fig. \ref{fig.a2_diam}). While this method of estimation is very crude, the exact numbers are not overly important, as we are only interested in using the scaling relation to determine rough Yarkovsky detectability thresholds for undiscovered asteroids. 

\begin{figure*}
\centering
\includegraphics[width=2\columnwidth]{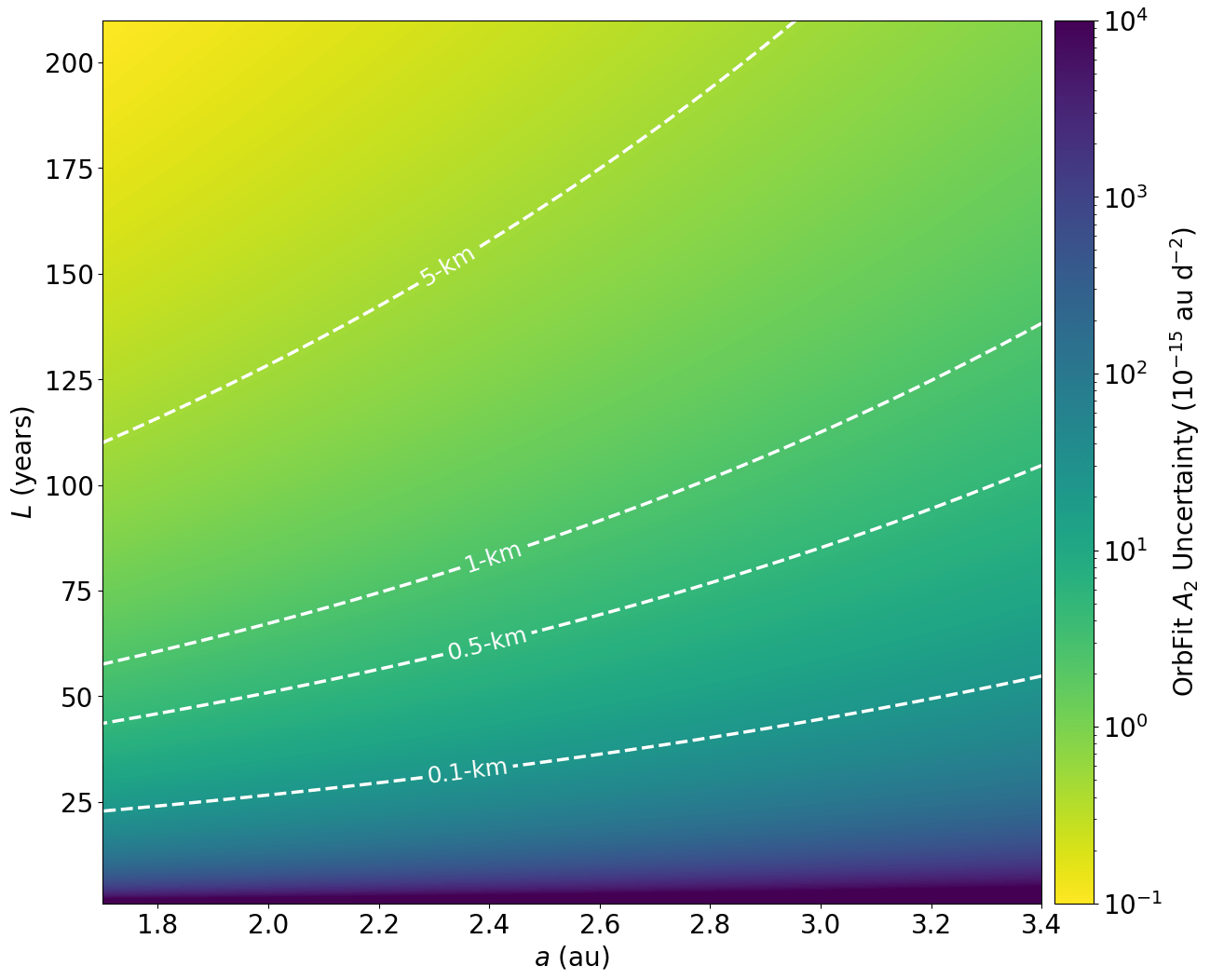}
\caption{Equation \ref{eq.2dfit} is represented here as a three-dimensional plot, with the $A_2$ uncertainty as a function of the semimajor axis $a$ and observational arc length $L$. Overplotted are contour lines for the minimum arc length for the Yarkovsky effect to become detectable as a function of $a$ for MBAs of 0.1, 0.5, 1, and 5 km in diameter, where their $A_2$ values are estimated from the diameter scaling relation described in Figure \ref{fig.a2_diam}.}
\label{fig.fit_3d}
\end{figure*}

Using the modified scaling relation, we expect a 100 m MBA to have an $A_{2, \mathrm{exp}}$ of $73 \times 10^{-15}$ au day$^{-2}$ and a 1-km MBA to have an $A_{2, \mathrm{exp}}$ of $7.3 \times 10^{-15}$ au day$^{-2}$. With Equation \ref{eq.2dfit}, we can find the minimum arc length to detect the Yarkovsky effect as a function of semimajor axis. While with current technology, it would only become possible to detect a Yarkovsky drift in the central main belt region in about 80 yr for a 1 km MBA, detection is dramatically easier for 100 m MBAs, where it would take about 30 yr (Fig. \ref{fig.fit_3d}). So long as we can find such small MBAs and maintain consistent high-quality astrometry, a Yarkovsky detection in the main belt is achievable within a reasonable time frame.

\section{Conclusions and Future Prospects}

Although there are hundreds of reported cases among the NEA population, the Yarkovsky effect has never been detected in individual MBAs, and that fact remains true with our attempts in this work. Using our sample of over a thousand thermophysically modeled MBAs, we were able to identify among them the most promising candidates for direct Yarkovsky detection based on their predicted $A_2$ values. We observed a subset of these candidates with the UH88, and we attempted to look for direct signs of Yarkovsky acceleration with the JPL asteroid and comet orbit determination software, where our observational data consisted of data from our UH88 observations, Gaia DR3, and archival astrometry reported at the MPC.

The majority of MBAs in our sample have archival data spanning many decades, with a few having total arc lengths of longer than a century. Long observational arcs provide greater constraints for the orbit determination and thus are advantageous to have when attempting to detect Yarkovsky-induced orbital drift. However, the historical data from up through the mid-20th century are often of much cruder precision and lower quality in comparison to astrometric positions from the past few decades \citep{Veres17} thanks to technological advancements across telescopes and detectors \citet[see e.g., the review by][]{Tokunaga14} as well as larger and more accurate star catalogs \citep[e.g.,][]{Anders22, Gaia22}. With the release of Gaia DR3 \citep{Tanga22}, we have hundreds of yet-unused highly accurate astrometric positions for many of the asteroids in our sample. As expected, the improvements to the orbit solutions provided by these observations were largely marginal and not sufficient to produce a plausible Yarkovsky detection among any of our MBAs. The Gaia mission was only launched in 2013, in the modern era where observations of MBAs from ground-based observatories are plentiful and a regular occurrence.

We showed that the existing observations are not of sufficiently high accuracy or precision to constrain the Yarkovsky effect. Using synthetic observations assuming a certain value for $A_2$, we found the minimum observational arc length required to detect the Yarkovsky effect given an expected $A_2$ and semimajor axis. It should be stressed, however, that this minimum is under the best-case conditions, where the asteroid is regularly observed throughout the whole arc and at a level of accuracy (0.$\arcsec$1) that is generally high even for modern-day astrometry \citep{Veres17}. Note that we also do not take into consideration the viability of an individual observation (e.g., limitations on magnitude or solar elongation), but main belt orbits are easily accessible for observations on a yearly basis, and the interval between observations is not significant as long as it is short compared with the total arc length. Searches for Yarkovsky acceleration have sometimes found detections that are strongly dependent on a few observations that are separated by a long time interval from the rest of the data set \citep[e.g.,][]{Farnocchia13, DelVigna18}. Remeasuring these isolated observations against modern star catalogs could confirm whether or not the Yarkovsky acceleration signal is real, as was done for the four 1953 precovery observations of (152563) 1992\,BF \citep{Vokrouhlicky08}, but the data may not necessarily have been preserved to make such remeasurements possible.

Even if we were to find a valid Yarkovsky detection in the main belt, it cannot be considered reliable without the full consideration of the effects of the completeness of the perturbing asteroids and the uncertainties in their masses. While NEAs are relatively isolated, MBAs will spend more time in closer proximity to other asteroids in the main belt. Because of this, MBAs may be more sensitive to the gravitational perturbations of other asteroids than NEAs. The effects of perturbing asteroids may be inaccurately modeled, as methods for asteroid mass determination are limited, and derived masses are often associated with large uncertainties \citep[e.g.,][]{Carry12}. The most precise mass estimates come from spacecraft flybys of asteroids, but such missions are incredibly costly and thus very few in number \citep[e.g.,][]{Patzold11, Konopliv14, Lauretta17}. Multiple-asteroid systems provide the next most reliable method for mass determination through Kepler's third law and knowledge of the satellites' orbits around the primary asteroid \citep[e.g.,][]{Fang11, Rojo11, Yang20, Drummond21, Vernazza21}. The most common mass estimates have come from close encounters with other asteroids \citep[e.g.,][]{Michalak00, Michalak01, Kovacevic12, Baer17, Li19} or Mars \citep[e.g.,][]{Pitjeva01, Mouret09}, but the accuracy can be very crude for smaller asteroids \citep[e.g.,][]{Zielenbach11, Carry12}. Mass estimates are only available for a few hundred asteroids \citep{Carry12}; thus, it is possible that significant perturbers are not included in the dynamical models at all. If such an unknown perturber happened to be on an orbit trailing or leading a target asteroid, its gravitational contribution might end up mimicking or masking the effects of Yarkovsky acceleration. The perturber sensitivity problem was explored in depth with the Yarkovsky detection of Bennu \citep{Farnocchia21}, where the model included a total of 343 asteroid perturbers with their mass uncertainties, and the convergence of the orbital solution indicated that the perturber set was sufficient.

The best path forward to confirm the Yarkovsky effect in the main belt is simply more time, but the future is promising. Many MBAs are bright enough to be easily seen by survey telescopes such as the Panoramic Survey Telescope and Rapid Response System (Pan-STARRS1) telescope \citep{panstarrs} and the Catalina Sky Survey. These surveys discover thousands of new NEAs every year by scanning the entire night sky on a routine basis\footnote{\url{https://cneos.jpl.nasa.gov/stats/}} and at the same time obtain many incidental follow-up observations of MBAs, which have orbits ensuring that they will be observable at regular, roughly yearly intervals. Upcoming near-future surveys such as LSST \citep{lsst} are especially promising, as they will be capable of reaching deeper magnitudes and discovering much smaller MBAs where the Yarkovsky force is stronger and more easily detected. With our synthetic observations, we have shown that the Yarkovsky acceleration should be detectable in 100 m MBAs within a couple of decades. LSST in particular will be especially fruitful for asteroid discovery and follow-up, as it will operate in the Southern Hemisphere, which has traditionally seen much less observational coverage than the Northern Hemisphere.

Although we have no direct confirmation of Yarkovsky-induced orbital drift on any individual asteroid in the main belt as of yet, detection is in a sense inevitable as the body of high-quality astrometry grows. There are no benchmark tests to verify the accuracy of thermophysically derived parameters, and ground-truth information is highly limited, restricted to spacecraft missions that are only feasible for a very small number of asteroids. However, a successful detection of Yarkovsky acceleration could serve as a means of constraining the physical properties of an asteroid that are independent of its thermal data and the limitations of the thermophysical models. The first MBA Yarkovsky detection may end up occurring with a yet-undiscovered asteroid, but thermal data of asteroids will only grow more plentiful with upcoming space-based surveys like the Near-Earth Object Surveyor\footnote{\url{https://neos.arizona.edu/}}. The detection of Bennu's Yarkovsky acceleration combined with its thermophysically derived parameters made it possible to estimate its bulk density \citep{Chesley14}, a value that was later confirmed by spacecraft tracking measurements \citep{Scheeres19}. Of course, these are only hypothetical scenarios for MBAs for now, but in time, we will have a better understanding of the makeup of the largest group of asteroids in the solar system.

\section*{Acknowledgements}

{\footnotesize We thank the two anonymous referees for their valuable comments in improving this work. The work of D.H. and D.J.T. are supported by NASA grant Nos. NNX13AI64G and 80NSSC21K0807. Part of this work was carried out at the Jet Propulsion Laboratory, California Institute of Technology, under a contract with the National Aeronautics and Space Administration (80NM0018D0004). This research has made use of data and/or services provided by the International Astronomical Union's Minor Planet Center. This publication makes use of data products from the Wide-field Infrared Survey Explorer, which is a joint project of the University of California, Los Angeles, and the Jet Propulsion Laboratory/California Institute of Technology, funded by the National Aeronautics and Space Administration. This publication also makes use of data products from NEOWISE, which is a project of the Jet Propulsion Laboratory/California Institute of Technology, funded by the Planetary Science Division of the National Aeronautics and Space Administration. We thank the indigenous Hawaiian community for allowing us to be guests on their sacred mountain, a privilege, without which, this work would not have been possible. We are most fortunate to be able to conduct observations from this site.}

\bibliographystyle{aasjournal}
\bibliography{yarkthermo}

\section{Appendix}

\startlongtable
\begin{deluxetable*}{rlccccccr}
\tablecolumns{9}
\tablecaption{UH88 Observations\label{tab.obs2}}
\tablehead{\twocolhead{Designation} & \colhead{2021} & \colhead{2021} & \colhead{2021} & \colhead{2021} & \colhead{2021} & \colhead{2022} & \colhead{Predicted $A_2$} \\
           \twocolhead{} & \colhead{Apr 13} & \colhead{May 12} & \colhead{Jun 10} & \colhead{Sep 25} & \colhead{Oct 23} & \colhead{Feb 27} & \colhead{($10^{15}$ au day$^{-2}$)} }
\startdata
767 & Bondia & \nodata & \nodata & \nodata & \nodata & \nodata & \checkmark & $0.07^{+0.01}_{-0.02}$ \\ 
769 & Tatjana & \nodata & \nodata & \nodata & \nodata & \nodata & \checkmark & $0.16^{+0.03}_{-0.16}$ \\ 
963 & Iduberga & \checkmark & \checkmark & \checkmark & \nodata & \nodata & \nodata & $0.93^{+0.10}_{-0.26}$ \\ 
1169 & Alwine & \checkmark & \checkmark & \checkmark & \nodata & \nodata & \nodata & $-0.91^{+0.64}_{-0.54}$ \\ 
1415 & Malautra & \nodata & \checkmark & \checkmark & \nodata & \nodata & \nodata & $0.25^{+0.10}_{-0.16}$ \\ 
1430 & Somalia & \checkmark & \checkmark & \nodata & \nodata & \nodata & \nodata & $0.70^{+0.16}_{-0.70}$ \\ 
1518 & Rovaniemi & \nodata & \nodata & \checkmark & \nodata & \nodata & \nodata & $0.91^{+0.16}_{-0.37}$ \\ 
1652 & Herge & \nodata & \nodata & \checkmark & \nodata & \nodata & \nodata & $0.67^{+0.02}_{-0.04}$ \\ 
1685 & Toro & \checkmark & \checkmark & \checkmark & \nodata & \nodata & \nodata & $-2.60^{+0.31}_{-0.19}$ \\ 
1704 & Wachmann & \nodata & \checkmark & \checkmark & \nodata & \nodata & \nodata & $0.94^{+0.14}_{-0.20}$ \\ 
1772 & Gagarin & \nodata & \checkmark & \nodata & \nodata & \nodata & \nodata & $0.05^{+0.02}_{-0.01}$ \\ 
1773 & Rumpelstilz & \checkmark & \checkmark & \checkmark & \nodata & \nodata & \nodata & $0.77^{+0.05}_{-0.09}$ \\ 
1865 & Cerberus & \nodata & \nodata & \nodata & \checkmark & \checkmark & \nodata & $-7.18^{+0.85}_{-0.57}$ \\ 
2204 & Lyyli & \nodata & \nodata & \nodata & \nodata & \nodata & \checkmark & $-0.46^{+0.10}_{-0.02}$ \\ 
2493 & Elmer & \checkmark & \checkmark & \nodata & \nodata & \nodata & \nodata & $-0.77^{+0.69}_{-0.23}$ \\ 
2818 & Juvenalis & \nodata & \nodata & \nodata & \checkmark & \nodata & \nodata & $-2.22^{+2.22}_{-0.62}$ \\ 
2840 & Kallavesi & \checkmark & \checkmark & \checkmark & \nodata & \nodata & \nodata & $0.00^{+0.41}_{-0.00}$ \\ 
2874 & Jim Young & \checkmark & \checkmark & \checkmark & \nodata & \nodata & \nodata & $0.71^{+0.65}_{-0.71}$ \\ 
3121 & Tamines & \checkmark & \checkmark & \checkmark & \nodata & \nodata & \nodata & $0.27^{+0.48}_{-0.17}$ \\ 
3376 & Armandhammer & \checkmark & \checkmark & \checkmark & \nodata & \nodata & \nodata & $0.11^{+0.04}_{-0.02}$ \\ 
3383 & Koyama & \checkmark & \checkmark & \checkmark & \nodata & \nodata & \nodata & $0.81^{+0.18}_{-0.25}$ \\ 
3510 & Veeder & \checkmark & \checkmark & \checkmark & \nodata & \nodata & \nodata & $0.71^{+0.06}_{-0.13}$ \\ 
3556 & Lixiaohua & \nodata & \nodata & \nodata & \nodata & \nodata & \checkmark & $-0.00^{+0.00}_{-0.51}$ \\ 
3722 & Urata & \nodata & \nodata & \nodata & \nodata & \nodata & \checkmark & $-0.03^{+0.01}_{-0.16}$ \\ 
4005 & Dyagilev & \checkmark & \checkmark & \checkmark & \nodata & \nodata & \nodata & $0.28^{+0.11}_{-0.08}$ \\ 
4088 & Baggesen & \nodata & \checkmark & \checkmark & \nodata & \nodata & \nodata & $0.61^{+0.19}_{-0.27}$ \\ 
4113 & Rascana & \nodata & \nodata & \checkmark & \nodata & \nodata & \nodata & $-0.19^{+0.09}_{-0.48}$ \\ 
4285 & Hulkower & \nodata & \nodata & \nodata & \nodata & \nodata & \checkmark & $-1.28^{+1.28}_{-0.24}$ \\ 
4323 & Hortulus & \checkmark & \checkmark & \checkmark & \nodata & \nodata & \nodata & $1.31^{+0.34}_{-0.56}$ \\ 
4399 & Ashizuri & \nodata & \checkmark & \checkmark & \nodata & \nodata & \nodata & $-0.30^{+0.17}_{-0.57}$ \\ 
4515 & Khrennikov & \nodata & \checkmark & \checkmark & \nodata & \nodata & \nodata & $-1.60^{+1.03}_{-0.70}$ \\ 
4912 & Emilhaury & \checkmark & \checkmark & \checkmark & \nodata & \nodata & \nodata & $-1.46^{+1.14}_{-0.25}$ \\ 
4928 & Vermeer & \checkmark & \checkmark & \nodata & \nodata & \nodata & \nodata & $-1.99^{+0.64}_{-0.43}$ \\ 
4959 & Niinoama & \nodata & \nodata & \nodata & \nodata & \nodata & \checkmark & $-0.32^{+0.05}_{-0.04}$ \\ 
5589 & De Meis & \checkmark & \checkmark & \nodata & \nodata & \nodata & \nodata & $-1.15^{+0.96}_{-0.49}$ \\ 
5635 & Cole & \checkmark & \checkmark & \nodata & \nodata & \nodata & \nodata & $0.32^{+1.73}_{-0.14}$ \\ 
5889 & Mickiewicz & \nodata & \nodata & \nodata & \nodata & \nodata & \checkmark & $-0.70^{+0.15}_{-0.10}$ \\ 
5958 & Barrande & \nodata & \checkmark & \nodata & \nodata & \nodata & \nodata & $-0.05^{+0.02}_{-0.30}$ \\ 
6499 & Michiko & \nodata & \checkmark & \nodata & \nodata & \nodata & \nodata & $-0.74^{+0.49}_{-0.43}$ \\ 
6581 & Sobers & \nodata & \nodata & \nodata & \nodata & \nodata & \checkmark & $-1.99^{+1.21}_{-0.35}$ \\ 
6607 & Matsushima & \nodata & \nodata & \nodata & \nodata & \nodata & \checkmark & $-0.71^{+0.71}_{-0.15}$ \\ 
6714 & Montreal & \checkmark & \checkmark & \checkmark & \nodata & \nodata & \nodata & $0.05^{+0.51}_{-0.02}$ \\ 
6755 & Solov'yanenko & \nodata & \checkmark & \checkmark & \nodata & \nodata & \nodata & $1.17^{+0.22}_{-0.69}$ \\ 
6915 & (1992\,HH) & \checkmark & \checkmark & \checkmark & \nodata & \nodata & \nodata & $-0.51^{+0.25}_{-0.35}$ \\ 
7196 & Baroni & \nodata & \nodata & \nodata & \checkmark & \checkmark & \nodata & $-2.17^{+1.15}_{-0.54}$ \\ 
7517 & Alisondoane & \checkmark & \checkmark & \checkmark & \nodata & \nodata & \nodata & $1.04^{+0.07}_{-0.22}$ \\ 
7684 & Marioferrero & \checkmark & \checkmark & \checkmark & \nodata & \nodata & \nodata & $1.06^{+0.27}_{-0.47}$ \\ 
8359 & (1989\,WD) & \nodata & \nodata & \checkmark & \nodata & \nodata & \nodata & $-1.85^{+0.67}_{-0.79}$ \\ 
9008 & Bohsternberk & \nodata & \nodata & \checkmark & \nodata & \nodata & \nodata & $-1.73^{+0.40}_{-0.20}$ \\ 
9158 & Plate & \checkmark & \checkmark & \nodata & \nodata & \nodata & \nodata & $-1.55^{+0.76}_{-0.44}$ \\ 
9173 & Viola Castello & \nodata & \checkmark & \nodata & \nodata & \nodata & \nodata & $-1.15^{+1.15}_{-0.24}$ \\ 
9234 & Matsumototaku & \nodata & \nodata & \nodata & \checkmark & \checkmark & \nodata & $-2.67^{+2.56}_{-0.74}$ \\ 
9274 & Amylovell & \nodata & \nodata & \nodata & \nodata & \checkmark & \nodata & $-2.63^{+1.28}_{-1.14}$ \\ 
9566 & Rykhlova & \nodata & \nodata & \nodata & \nodata & \checkmark & \nodata & $-1.97^{+0.55}_{-0.23}$ \\ 
9582 & (1990\,EL7) & \nodata & \nodata & \nodata & \nodata & \checkmark & \nodata & $-2.00^{+2.00}_{-0.48}$ \\ 
10166 & Takarajima & \checkmark & \checkmark & \checkmark & \nodata & \nodata & \nodata & $-0.00^{+0.00}_{-2.48}$ \\ 
10338 & (1991\,RB11) & \nodata & \nodata & \nodata & \nodata & \nodata & \checkmark & $-0.09^{+0.08}_{-0.78}$ \\ 
10406 & (1997\,WZ29) & \nodata & \nodata & \nodata & \nodata & \nodata & \checkmark & $-1.35^{+0.32}_{-0.17}$ \\ 
10456 & Anechka & \checkmark & \checkmark & \checkmark & \nodata & \nodata & \nodata & $1.82^{+0.36}_{-0.79}$ \\ 
10656 & Albrecht & \checkmark & \checkmark & \nodata & \nodata & \nodata & \nodata & $-0.61^{+0.45}_{-0.14}$ \\ 
11676 & (1998\,CQ2) & \nodata & \nodata & \nodata & \checkmark & \checkmark & \nodata & $-2.64^{+0.34}_{-0.30}$ \\ 
11823 & Christen & \nodata & \nodata & \nodata & \nodata & \nodata & \checkmark & $1.77^{+0.31}_{-1.44}$ \\ 
11889 & (1991\,AH2) & \checkmark & \checkmark & \checkmark & \nodata & \nodata & \nodata & $0.46^{+0.04}_{-0.30}$ \\ 
12097 & (1998\,HG121) & \nodata & \checkmark & \checkmark & \nodata & \nodata & \nodata & $1.04^{+0.44}_{-0.40}$ \\ 
12374 & Rakhat & \checkmark & \checkmark & \nodata & \nodata & \nodata & \nodata & $1.72^{+0.36}_{-0.70}$ \\ 
12555 & (1998\,QP47) & \checkmark & \checkmark & \checkmark & \nodata & \nodata & \nodata & $-0.58^{+0.58}_{-0.51}$ \\ 
12617 & Angelusilesius & \nodata & \nodata & \nodata & \nodata & \nodata & \checkmark & $-0.00^{+0.00}_{-0.99}$ \\ 
12690 & Kochimiraikagaku & \nodata & \nodata & \nodata & \nodata & \nodata & \checkmark & $-0.21^{+0.21}_{-0.14}$ \\ 
12705 & (1990\,TJ) & \checkmark & \checkmark & \checkmark & \nodata & \nodata & \nodata & $-0.51^{+0.31}_{-0.41}$ \\ 
12883 & (1998\,QY32) & \checkmark & \checkmark & \checkmark & \nodata & \nodata & \nodata & $-0.00^{+0.00}_{-0.64}$ \\ 
12926 & Brianmason & \nodata & \nodata & \nodata & \checkmark & \checkmark & \nodata & $2.06^{+0.37}_{-0.52}$ \\ 
13007 & (1984\,AU) & \checkmark & \checkmark & \checkmark & \nodata & \nodata & \nodata & $-0.65^{+0.65}_{-0.67}$ \\ 
13019 & (1988\,NW) & \checkmark & \checkmark & \nodata & \nodata & \nodata & \nodata & $0.71^{+0.35}_{-0.42}$ \\ 
13059 & Ducuroir & \checkmark & \checkmark & \checkmark & \nodata & \nodata & \nodata & $1.89^{+0.34}_{-0.60}$ \\ 
13446 & Almarkim & \nodata & \nodata & \nodata & \nodata & \nodata & \checkmark & $0.00^{+0.07}_{-0.00}$ \\ 
13883 & (7066\,P-L) & \nodata & \nodata & \nodata & \nodata & \nodata & \checkmark & $1.01^{+0.09}_{-0.16}$ \\ 
13925 & (1986\,QS3) & \checkmark & \checkmark & \checkmark & \nodata & \nodata & \nodata & $-0.57^{+0.16}_{-0.14}$ \\ 
13936 & (1989\,HC) & \nodata & \nodata & \nodata & \nodata & \nodata & \checkmark & $0.95^{+0.07}_{-0.25}$ \\ 
13948 & (1990\,QB6) & \nodata & \nodata & \nodata & \nodata & \nodata & \checkmark & $0.43^{+0.63}_{-0.24}$ \\ 
14031 & Rozyo & \checkmark & \checkmark & \checkmark & \nodata & \nodata & \nodata & $-0.88^{+0.88}_{-0.46}$ \\ 
14257 & (2000\,AR97) & \nodata & \nodata & \checkmark & \nodata & \nodata & \nodata & $0.62^{+0.94}_{-0.62}$ \\ 
15920 & (1997\,UB25) & \nodata & \nodata & \nodata & \nodata & \checkmark & \nodata & $-2.40^{+1.72}_{-0.49}$ \\ 
17431 & Sainte-Colombe & \nodata & \nodata & \nodata & \checkmark & \checkmark & \nodata & $2.12^{+0.38}_{-0.53}$ \\ 
18591 & (1997\,YT11) & \nodata & \nodata & \checkmark & \nodata & \nodata & \nodata & $0.48^{+0.72}_{-0.34}$ \\ 
18997 & Mizrahi & \nodata & \nodata & \nodata & \checkmark & \checkmark & \nodata & $-2.66^{+2.66}_{-0.46}$ \\ 
19136 & Strassmann & \nodata & \nodata & \nodata & \checkmark & \checkmark & \nodata & $-2.27^{+2.27}_{-0.48}$ \\ 
19876 & (7637\,P-L) & \nodata & \nodata & \nodata & \checkmark & \checkmark & \nodata & $2.09^{+0.32}_{-1.89}$ \\ 
20179 & (1996\,XX31) & \nodata & \nodata & \nodata & \checkmark & \checkmark & \nodata & $-2.36^{+1.30}_{-0.58}$ \\ 
20498 & (1999\,RT1) & \nodata & \nodata & \nodata & \nodata & \nodata & \checkmark & $0.00^{+1.08}_{-0.00}$ \\ 
20515 & (1999\,RO34) & \nodata & \nodata & \nodata & \nodata & \nodata & \checkmark & $0.91^{+0.29}_{-0.39}$ \\ 
20557 & Davidkulka & \checkmark & \checkmark & \checkmark & \nodata & \nodata & \nodata & $1.42^{+0.29}_{-0.60}$ \\ 
20681 & (1999\,VH10) & \nodata & \checkmark & \checkmark & \nodata & \nodata & \nodata & $0.99^{+0.27}_{-0.56}$ \\ 
21041 & (1990\,QO1) & \nodata & \nodata & \nodata & \checkmark & \checkmark & \nodata & $-2.06^{+1.29}_{-0.39}$ \\ 
21842 & (1999\,TH102) & \nodata & \nodata & \nodata & \nodata & \nodata & \checkmark & $-0.44^{+0.44}_{-0.25}$ \\ 
22092 & (2000\,AQ199) & \nodata & \nodata & \nodata & \nodata & \nodata & \checkmark & $-0.00^{+0.00}_{-0.23}$ \\ 
24181 & (1999\,XN8) & \checkmark & \checkmark & \checkmark & \nodata & \nodata & \nodata & $-1.38^{+1.38}_{-0.55}$ \\ 
25887 & (2000\,SU308) & \nodata & \nodata & \nodata & \checkmark & \checkmark & \nodata & $2.34^{+0.32}_{-0.86}$ \\ 
26176 & (1996\,GD2) & \nodata & \nodata & \nodata & \checkmark & \checkmark & \nodata & $-2.22^{+0.59}_{-0.58}$ \\ 
26520 & (2000\,CQ75) & \nodata & \nodata & \nodata & \nodata & \nodata & \checkmark & $1.04^{+0.34}_{-1.04}$ \\ 
27259 & (1999\,XS136) & \checkmark & \checkmark & \nodata & \nodata & \nodata & \nodata & $1.18^{+0.11}_{-0.14}$ \\ 
28709 & (2000\,GY96) & \nodata & \nodata & \nodata & \nodata & \nodata & \checkmark & $-1.41^{+0.28}_{-0.15}$ \\ 
28736 & (2000\,GE133) & \nodata & \nodata & \nodata & \checkmark & \checkmark & \nodata & $-2.09^{+0.49}_{-0.18}$ \\ 
30072 & (2000\,EP93) & \checkmark & \checkmark & \nodata & \nodata & \nodata & \nodata & $0.90^{+0.26}_{-0.54}$ \\ 
31755 & (1999\,JA96) & \nodata & \nodata & \nodata & \nodata & \nodata & \checkmark & $-0.95^{+0.95}_{-0.92}$ \\ 
31829 & (1999\,XT12) & \nodata & \nodata & \nodata & \nodata & \nodata & \checkmark & $-0.59^{+0.59}_{-0.89}$ \\ 
32103 & Re'emsari & \nodata & \nodata & \nodata & \nodata & \nodata & \checkmark & $-1.55^{+1.22}_{-0.88}$ \\ 
32507 & (2001\,LR15) & \nodata & \nodata & \nodata & \checkmark & \checkmark & \nodata & $-2.88^{+0.71}_{-0.56}$ \\ 
32776 & Nriag & \checkmark & \checkmark & \checkmark & \nodata & \nodata & \nodata & $-1.05^{+0.66}_{-0.92}$ \\ 
33116 & (1998\,BO12) & \nodata & \nodata & \checkmark & \nodata & \nodata & \nodata & $1.52^{+0.43}_{-0.46}$ \\ 
33181 & Aalokpatwa & \nodata & \nodata & \nodata & \checkmark & \nodata & \nodata & $-2.30^{+0.54}_{-0.29}$ \\ 
33974 & (2000\,ND17) & \nodata & \nodata & \nodata & \nodata & \nodata & \checkmark & $-2.24^{+1.27}_{-0.59}$ \\ 
34290 & (2000\,QQ150) & \nodata & \nodata & \nodata & \nodata & \nodata & \checkmark & $0.02^{+0.07}_{-0.02}$ \\ 
35595 & (1998\,HO116) & \nodata & \nodata & \nodata & \nodata & \nodata & \checkmark & $-0.24^{+0.24}_{-0.27}$ \\ 
37377 & (2001\,VP46) & \nodata & \checkmark & \checkmark & \nodata & \nodata & \nodata & $-0.08^{+0.03}_{-0.05}$ \\ 
38650 & (2000\,ON17) & \nodata & \nodata & \nodata & \checkmark & \checkmark & \nodata & $1.70^{+0.54}_{-0.57}$ \\ 
38950 & (2000\,ST295) & \nodata & \nodata & \nodata & \checkmark & \checkmark & \nodata & $-2.39^{+0.79}_{-0.48}$ \\ 
40104 & (1998\,QE4) & \nodata & \nodata & \nodata & \nodata & \nodata & \checkmark & $-1.16^{+0.20}_{-0.41}$ \\ 
40413 & (1999\,RS10) & \checkmark & \checkmark & \checkmark & \nodata & \nodata & \nodata & $1.64^{+0.51}_{-1.26}$ \\ 
41709 & (2000\,UH56) & \checkmark & \checkmark & \checkmark & \nodata & \nodata & \nodata & $1.21^{+0.79}_{-0.51}$ \\ 
42490 & (1991\,SU) & \nodata & \nodata & \nodata & \checkmark & \checkmark & \nodata & $1.80^{+0.22}_{-1.80}$ \\ 
44612 & (1999\,RP27) & \checkmark & \checkmark & \nodata & \nodata & \nodata & \nodata & $-2.20^{+1.46}_{-2.61}$ \\ 
45898 & (2000\,XQ49) & \nodata & \nodata & \nodata & \checkmark & \checkmark & \nodata & $2.53^{+0.86}_{-1.12}$ \\ 
47127 & (1999\,CJ103) & \checkmark & \checkmark & \nodata & \nodata & \nodata & \nodata & $-0.83^{+0.35}_{-0.29}$ \\ 
49088 & (1998\,RS68) & \checkmark & \checkmark & \checkmark & \nodata & \nodata & \nodata & $-1.99^{+0.42}_{-0.22}$ \\ 
50219 & (2000\,AL237) & \checkmark & \checkmark & \checkmark & \nodata & \nodata & \nodata & $-0.16^{+0.05}_{-0.76}$ \\ 
50776 & (2000\,FS12) & \checkmark & \checkmark & \nodata & \nodata & \nodata & \nodata & $0.75^{+0.49}_{-0.44}$ \\ 
51291 & (2000\,KK29) & \nodata & \nodata & \nodata & \nodata & \nodata & \checkmark & $-0.00^{+0.00}_{-0.35}$ \\ 
51822 & (2001\,OB25) & \nodata & \nodata & \nodata & \nodata & \nodata & \checkmark & $-0.26^{+0.16}_{-1.63}$ \\ 
56131 & (1999\,CY48) & \checkmark & \checkmark & \checkmark & \nodata & \nodata & \nodata & $1.05^{+1.96}_{-1.05}$ \\ 
57429 & (2001\,SX33) & \nodata & \nodata & \nodata & \nodata & \checkmark & \nodata & $-1.94^{+1.94}_{-0.56}$ \\ 
59150 & (1998\,XV90) & \nodata & \nodata & \nodata & \checkmark & \checkmark & \nodata & $-2.06^{+0.44}_{-0.28}$ \\ 
66003 & (1998\,OX6) & \checkmark & \checkmark & \nodata & \nodata & \nodata & \nodata & $-0.17^{+0.07}_{-0.96}$ \\ 
79056 & (1132\,T-3) & \nodata & \checkmark & \checkmark & \nodata & \checkmark & \nodata & $4.10^{+0.46}_{-1.52}$ \\ 
87932 & (2000\,SW343) & \nodata & \checkmark & \checkmark & \nodata & \nodata & \nodata & $1.05^{+0.17}_{-0.38}$ \\ 
89932 & (2002\,EV85) & \nodata & \nodata & \nodata & \checkmark & \checkmark & \nodata & $2.14^{+0.57}_{-1.87}$ \\ 
92930 & (2000\,RH26) & \nodata & \checkmark & \checkmark & \nodata & \nodata & \nodata & $-0.03^{+0.01}_{-0.06}$ \\ 
99028 & (2001\,DC98) & \nodata & \nodata & \nodata & \checkmark & \checkmark & \nodata & $-3.79^{+2.39}_{-0.65}$ \\ 
\vspace{-0.3cm}
\enddata
\tablecomments{Observations for each target asteroid separated by date. We observed a total of 134 unique MBAs, as well as one Mars-crosser (2204 Lyyli) and two NEAs (1685 Toro and 1865 Cerberus). Each night consists of at least three dithered exposures per target. Due to the use of multiple shape models in the thermophysical modeling, there is often more than one set of derived thermal properties for an asteroid. The predicted $A_2$ reported here uses the thermal inertia, diameter, and albedo derived from the thermophysical modeling fit with the smallest $\chi^2$ for that asteroid. Refer to \citet{Hung22} for more details. }
\end{deluxetable*}
\clearpage

\startlongtable
\begin{deluxetable*}{rlrccc}
\tablecolumns{7}
\tablecaption{$A_2$ Values from Orbit Determination\label{tab.a2s2}}
\tablehead{\twocolhead{Designation} &  \colhead{Predicted $A_2$} & \colhead{Derived $A_2$} & \colhead{S/N} & \colhead{$\mathcal{S}$} \\
           \twocolhead{} & \colhead{($10^{15}$ au day$^{-2}$)} & \colhead{($10^{15}$ au day$^{-2}$)} & \colhead{} & \colhead{}}
\startdata
   1685 & Toro             & $-2.60^{+0.31}_{-0.19}$ & $  -2.95 \pm   0.46$ & 6.40 &    1.1 \\ 
   1865 & Cerberus         & $-7.18^{+0.85}_{-0.57}$ & $ -11.90 \pm   3.69$ & 3.22 &    1.7 \\ 
   1773 & Rumpelstilz      & $ 0.77^{+0.05}_{-0.09}$ & $ -67.23 \pm  26.41$ & 2.55 &   87.2 \\ 
  22092 & 2000\,AQ199      & $-0.00^{+0.00}_{-0.23}$ & $-132.11 \pm  52.99$ & 2.49 &  575.1 \\ 
   1430 & Somalia          & $ 0.70^{+0.16}_{-0.70}$ & $ -63.42 \pm  29.32$ & 2.16 &   90.3 \\ 
   9158 & Plate            & $-1.55^{+0.76}_{-0.44}$ & $ -94.07 \pm  46.56$ & 2.02 &   60.6 \\ 
  99028 & 2001\,DC98       & $-3.79^{+2.39}_{-0.65}$ & $-222.60 \pm 112.42$ & 1.98 &   58.8 \\ 
  12617 & Angelusilesius   & $-0.00^{+0.00}_{-0.99}$ & $  91.37 \pm  46.30$ & 1.97 &   92.3 \\ 
    769 & Tatjana          & $ 0.16^{+0.03}_{-0.16}$ & $-109.06 \pm  57.43$ & 1.90 &  662.2 \\ 
  15920 & 1997\,UB25       & $-2.40^{+1.72}_{-0.49}$ & $ -76.05 \pm  40.66$ & 1.87 &   31.7 \\ 
   4005 & Dyagilev         & $ 0.28^{+0.11}_{-0.08}$ & $  66.40 \pm  35.53$ & 1.87 &  236.4 \\ 
  10166 & Takarajima       & $-0.00^{+0.00}_{-2.48}$ & $-157.51 \pm  85.72$ & 1.84 &   63.6 \\ 
   6714 & Montreal         & $ 0.05^{+0.51}_{-0.02}$ & $ -93.12 \pm  52.11$ & 1.79 & 1808.1 \\ 
   2493 & Elmer            & $-0.77^{+0.69}_{-0.23}$ & $ -99.08 \pm  56.05$ & 1.77 &  128.4 \\ 
  56131 & 1999\,CY48       & $ 1.05^{+1.96}_{-1.05}$ & $ -99.73 \pm  57.74$ & 1.73 &   95.1 \\ 
   3376 & Armandhammer     & $ 0.11^{+0.04}_{-0.02}$ & $ -65.37 \pm  38.82$ & 1.68 &  581.1 \\ 
  13007 & 1984\,AU         & $-0.65^{+0.65}_{-0.67}$ & $ -80.10 \pm  48.05$ & 1.67 &  124.1 \\ 
  11823 & Christen         & $ 1.77^{+0.31}_{-1.44}$ & $  60.24 \pm  36.05$ & 1.67 &   33.9 \\ 
  51822 & 2001\,OB25       & $-0.26^{+0.16}_{-1.63}$ & $ 247.81 \pm 149.33$ & 1.66 &  945.1 \\ 
  66003 & 1998\,OX6        & $-0.17^{+0.07}_{-0.96}$ & $-153.91 \pm  93.67$ & 1.64 &  918.8 \\ 
  21842 & 1999\,TH102      & $-0.44^{+0.44}_{-0.25}$ & $ 215.11 \pm 133.34$ & 1.61 &  491.0 \\ 
   6607 & Matsushima       & $-0.71^{+0.71}_{-0.15}$ & $ -94.93 \pm  59.41$ & 1.60 &  133.4 \\ 
  28709 & 2000\,GY96       & $-1.41^{+0.28}_{-0.15}$ & $-180.85 \pm 113.66$ & 1.59 &  128.6 \\ 
  14031 & Rozyo            & $-0.88^{+0.88}_{-0.46}$ & $ 109.93 \pm  70.44$ & 1.56 &  124.5 \\ 
   3383 & Koyama           & $ 0.81^{+0.18}_{-0.25}$ & $ -79.74 \pm  51.43$ & 1.55 &   98.1 \\ 
   1169 & Alwine           & $-0.91^{+0.64}_{-0.54}$ & $  30.82 \pm  19.89$ & 1.55 &   34.1 \\ 
   2204 & Lyyli            & $-0.46^{+0.10}_{-0.02}$ & $  51.40 \pm  34.80$ & 1.48 &  111.5 \\ 
  12555 & 1998\,QP47       & $-0.58^{+0.58}_{-0.51}$ & $-151.69 \pm 104.57$ & 1.45 &  260.4 \\ 
   6915 & 1992\,HH         & $-0.51^{+0.25}_{-0.35}$ & $ -87.24 \pm  60.51$ & 1.44 &  171.3 \\ 
  13948 & 1990\,QB6        & $ 0.43^{+0.63}_{-0.24}$ & $  67.89 \pm  48.30$ & 1.41 &  158.7 \\ 
  20557 & Davidkulka       & $ 1.42^{+0.29}_{-0.60}$ & $  56.03 \pm  41.45$ & 1.35 &   39.4 \\ 
   1704 & Wachmann         & $ 0.94^{+0.14}_{-0.20}$ & $ -24.57 \pm  18.49$ & 1.33 &   26.2 \\ 
  10456 & Anechka          & $ 1.82^{+0.36}_{-0.79}$ & $ -51.85 \pm  40.08$ & 1.29 &   28.5 \\ 
  12097 & 1998\,HG121      & $ 1.04^{+0.44}_{-0.40}$ & $ -49.86 \pm  38.89$ & 1.28 &   48.0 \\ 
   4959 & Niinoama         & $-0.32^{+0.05}_{-0.04}$ & $-115.12 \pm  89.65$ & 1.28 &  360.5 \\ 
  57429 & 2001\,SX33       & $-1.94^{+1.94}_{-0.56}$ & $-134.02 \pm 106.01$ & 1.26 &   69.0 \\ 
  13883 & 7066\,P-L        & $ 1.01^{+0.09}_{-0.16}$ & $ -96.52 \pm  77.20$ & 1.25 &   95.5 \\ 
  50219 & 2000\,AL237      & $-0.16^{+0.05}_{-0.76}$ & $ -73.76 \pm  61.54$ & 1.20 &  461.3 \\ 
  35595 & 1998\,HO116      & $-0.24^{+0.24}_{-0.27}$ & $ -86.04 \pm  76.66$ & 1.12 &  352.2 \\ 
  21041 & 1990\,QO1        & $-2.06^{+1.29}_{-0.39}$ & $ 130.52 \pm 121.29$ & 1.08 &   63.3 \\ 
   6581 & Sobers           & $-1.99^{+1.21}_{-0.35}$ & $  43.16 \pm  40.86$ & 1.06 &   21.7 \\ 
  32776 & Nriag            & $-1.05^{+0.66}_{-0.92}$ & $ -64.79 \pm  61.86$ & 1.05 &   61.9 \\ 
   4515 & Khrennikov       & $-1.60^{+1.03}_{-0.70}$ & $  30.36 \pm  29.02$ & 1.05 &   19.0 \\ 
  14257 & 2000\,AR97       & $ 0.62^{+0.94}_{-0.62}$ & $  32.12 \pm  30.90$ & 1.04 &   51.8 \\ 
  11676 & 1998\,CQ2        & $-2.64^{+0.34}_{-0.30}$ & $  55.80 \pm  57.68$ & 0.97 &   21.1 \\ 
   4912 & Emilhaury        & $-1.46^{+1.14}_{-0.25}$ & $ -29.45 \pm  30.95$ & 0.95 &   20.1 \\ 
   5635 & Cole             & $ 0.32^{+1.73}_{-0.14}$ & $ -42.96 \pm  46.89$ & 0.92 &  132.3 \\ 
  13936 & 1989\,HC         & $ 0.95^{+0.07}_{-0.25}$ & $-134.43 \pm 151.48$ & 0.89 &  140.8 \\ 
   7684 & Marioferrero     & $ 1.06^{+0.27}_{-0.47}$ & $ -62.41 \pm  69.84$ & 0.89 &   59.0 \\ 
   2818 & Juvenalis        & $-2.22^{+2.22}_{-0.62}$ & $ -26.81 \pm  30.19$ & 0.89 &   12.1 \\ 
  12883 & 1998\,QY32       & $-0.00^{+0.00}_{-0.64}$ & $ -34.51 \pm  39.71$ & 0.87 &   54.1 \\ 
  24181 & 1999\,XN8        & $-1.38^{+1.38}_{-0.55}$ & $ -66.27 \pm  77.33$ & 0.86 &   48.1 \\ 
  20515 & 1999\,RO34       & $ 0.91^{+0.29}_{-0.39}$ & $  91.40 \pm 112.53$ & 0.81 &  100.3 \\ 
  11889 & 1991\,AH2        & $ 0.46^{+0.04}_{-0.30}$ & $  62.90 \pm  78.07$ & 0.81 &  136.2 \\ 
  12374 & Rakhat           & $ 1.72^{+0.36}_{-0.70}$ & $ -29.69 \pm  38.88$ & 0.76 &   17.2 \\ 
  13925 & 1986\,QS3        & $-0.57^{+0.16}_{-0.14}$ & $  91.65 \pm 124.48$ & 0.74 &  159.8 \\ 
  42490 & 1991\,SU         & $ 1.80^{+0.22}_{-1.80}$ & $ -37.91 \pm  52.28$ & 0.73 &   21.0 \\ 
  12705 & 1990\,TJ         & $-0.51^{+0.31}_{-0.41}$ & $  32.20 \pm  44.43$ & 0.72 &   63.7 \\ 
  89932 & 2002\,EV85       & $ 2.14^{+0.57}_{-1.87}$ & $  65.15 \pm  91.31$ & 0.71 &   30.4 \\ 
  27259 & 1999\,XS136      & $ 1.18^{+0.11}_{-0.14}$ & $ -48.44 \pm  68.12$ & 0.71 &   41.0 \\ 
  13059 & Ducuroir         & $ 1.89^{+0.34}_{-0.60}$ & $ -42.03 \pm  59.17$ & 0.71 &   22.3 \\ 
  31755 & 1999\,JA96       & $-0.95^{+0.95}_{-0.92}$ & $ -70.41 \pm 100.50$ & 0.70 &   73.9 \\ 
   6499 & Michiko          & $-0.74^{+0.49}_{-0.43}$ & $ -60.84 \pm  87.48$ & 0.70 &   82.6 \\ 
  32507 & 2001\,LR15       & $-2.88^{+0.71}_{-0.56}$ & $  75.99 \pm 110.23$ & 0.69 &   26.4 \\ 
  13019 & 1988\,NW         & $ 0.71^{+0.35}_{-0.42}$ & $ -54.66 \pm  79.38$ & 0.69 &   77.1 \\ 
   3556 & Lixiaohua        & $-0.00^{+0.00}_{-0.51}$ & $ -58.95 \pm  85.06$ & 0.69 &  115.8 \\ 
  37377 & 2001\,VP46       & $-0.08^{+0.03}_{-0.05}$ & $ -34.95 \pm  51.40$ & 0.68 &  413.6 \\ 
  33116 & 1998\,BO12       & $ 1.52^{+0.43}_{-0.46}$ & $ -30.41 \pm  44.40$ & 0.68 &   20.0 \\ 
   5589 & De Meis          & $-1.15^{+0.96}_{-0.49}$ & $ -47.92 \pm  70.39$ & 0.68 &   41.7 \\ 
  31829 & 1999\,XT12       & $-0.59^{+0.59}_{-0.89}$ & $ -90.02 \pm 134.88$ & 0.67 &  153.1 \\ 
  30072 & 2000\,EP93       & $ 0.90^{+0.26}_{-0.54}$ & $ -63.06 \pm  95.34$ & 0.66 &   70.0 \\ 
   4113 & Rascana          & $-0.19^{+0.09}_{-0.48}$ & $ -19.54 \pm  29.77$ & 0.66 &  103.4 \\ 
  44612 & 1999\,RP27       & $-2.20^{+1.46}_{-2.61}$ & $  19.65 \pm  30.64$ & 0.64 &    8.9 \\ 
  26520 & 2000\,CQ75       & $ 1.04^{+0.34}_{-1.04}$ & $  38.77 \pm  62.25$ & 0.62 &   37.3 \\ 
  10338 & 1991\,RB11       & $-0.09^{+0.08}_{-0.78}$ & $  71.90 \pm 120.18$ & 0.60 &  805.1 \\ 
  10656 & Albrecht         & $-0.61^{+0.45}_{-0.14}$ & $  56.88 \pm  99.46$ & 0.57 &   92.5 \\ 
   3722 & Urata            & $-0.03^{+0.01}_{-0.16}$ & $ -12.76 \pm  22.54$ & 0.57 &  478.0 \\ 
   2840 & Kallavesi        & $ 0.00^{+0.41}_{-0.00}$ & $ -22.82 \pm  40.22$ & 0.57 &   56.1 \\ 
    963 & Iduberga         & $ 0.93^{+0.10}_{-0.26}$ & $  14.31 \pm  26.08$ & 0.55 &   15.4 \\ 
   1652 & Herge            & $ 0.67^{+0.02}_{-0.04}$ & $  -6.98 \pm  13.74$ & 0.51 &   10.4 \\ 
   8359 & 1989\,WD         & $-1.85^{+0.67}_{-0.79}$ & $  20.55 \pm  42.20$ & 0.49 &   11.1 \\ 
  25887 & 2000\,SU308      & $ 2.34^{+0.32}_{-0.86}$ & $ -74.54 \pm 154.08$ & 0.48 &   31.8 \\ 
   9234 & Matsumototaku    & $-2.67^{+2.56}_{-0.74}$ & $  15.27 \pm  33.32$ & 0.46 &    5.7 \\ 
  20681 & 1999\,VH10       & $ 0.99^{+0.27}_{-0.56}$ & $  33.58 \pm  74.61$ & 0.45 &   34.1 \\ 
   5889 & Mickiewicz       & $-0.70^{+0.15}_{-0.10}$ & $  44.89 \pm 103.54$ & 0.43 &   64.5 \\ 
  45898 & 2000\,XQ49       & $ 2.53^{+0.86}_{-1.12}$ & $ -10.45 \pm  25.52$ & 0.41 &    4.1 \\ 
  20179 & 1996\,XX31       & $-2.36^{+1.30}_{-0.58}$ & $ -19.15 \pm  46.71$ & 0.41 &    8.1 \\ 
  19136 & Strassmann       & $-2.27^{+2.27}_{-0.48}$ & $ -32.83 \pm  79.46$ & 0.41 &   14.5 \\ 
  12690 & Kochimiraikagaku & $-0.21^{+0.21}_{-0.14}$ & $ -36.87 \pm  94.55$ & 0.39 &  175.2 \\ 
   4399 & Ashizuri         & $-0.30^{+0.17}_{-0.57}$ & $ -20.70 \pm  53.75$ & 0.39 &   70.0 \\ 
  50776 & 2000\,FS12       & $ 0.75^{+0.49}_{-0.44}$ & $ -31.48 \pm  83.82$ & 0.38 &   42.0 \\ 
  40104 & 1998\,QE4        & $-1.16^{+0.20}_{-0.41}$ & $ -50.82 \pm 132.20$ & 0.38 &   43.8 \\ 
  92930 & 2000\,RH26       & $-0.03^{+0.01}_{-0.06}$ & $  49.98 \pm 134.54$ & 0.37 & 1566.8 \\ 
  18997 & Mizrahi          & $-2.66^{+2.66}_{-0.46}$ & $  27.47 \pm  74.70$ & 0.37 &   10.3 \\ 
   9173 & Viola Castello   & $-1.15^{+1.15}_{-0.24}$ & $ -27.95 \pm  80.43$ & 0.35 &   24.2 \\ 
    767 & Bondia           & $ 0.07^{+0.01}_{-0.02}$ & $  14.98 \pm  42.70$ & 0.35 &  229.8 \\ 
   3510 & Veeder           & $ 0.71^{+0.06}_{-0.13}$ & $  13.84 \pm  41.24$ & 0.34 &   19.5 \\ 
  47127 & 1999\,CJ103      & $-0.83^{+0.35}_{-0.29}$ & $  48.54 \pm 148.85$ & 0.33 &   58.2 \\ 
  26176 & 1996\,GD2        & $-2.22^{+0.59}_{-0.58}$ & $ -27.71 \pm  85.11$ & 0.33 &   12.5 \\ 
   1518 & Rovaniemi        & $ 0.91^{+0.16}_{-0.37}$ & $  -4.42 \pm  13.57$ & 0.33 &    4.9 \\ 
   9566 & Rykhlova         & $-1.97^{+0.55}_{-0.23}$ & $  14.02 \pm  43.25$ & 0.32 &    7.1 \\ 
   4928 & Vermeer          & $-1.99^{+0.64}_{-0.43}$ & $  -9.75 \pm  30.62$ & 0.32 &    4.9 \\ 
  28736 & 2000\,GE133      & $-2.09^{+0.49}_{-0.18}$ & $  11.25 \pm  36.83$ & 0.31 &    5.4 \\ 
   9008 & Bohsternberk     & $-1.73^{+0.40}_{-0.20}$ & $  -9.59 \pm  30.79$ & 0.31 &    5.5 \\ 
  19876 & 7637\,P-L        & $ 2.09^{+0.32}_{-1.89}$ & $  11.75 \pm  39.84$ & 0.30 &    5.6 \\ 
   1415 & Malautra         & $ 0.25^{+0.10}_{-0.16}$ & $   4.31 \pm  14.13$ & 0.30 &   17.2 \\ 
   4285 & Hulkower         & $-1.28^{+1.28}_{-0.24}$ & $ -15.14 \pm  55.78$ & 0.27 &   11.8 \\ 
  87932 & 2000\,SW343      & $ 1.05^{+0.17}_{-0.38}$ & $  23.27 \pm  91.76$ & 0.25 &   22.2 \\ 
  13446 & Almarkim         & $ 0.00^{+0.07}_{-0.00}$ & $  17.16 \pm  68.17$ & 0.25 &  260.4 \\ 
  40413 & 1999\,RS10       & $ 1.64^{+0.51}_{-1.26}$ & $ -12.75 \pm  52.18$ & 0.24 &    7.8 \\ 
   7517 & Alisondoane      & $ 1.04^{+0.07}_{-0.22}$ & $  -7.31 \pm  29.94$ & 0.24 &    7.1 \\ 
   4323 & Hortulus         & $ 1.31^{+0.34}_{-0.56}$ & $   5.35 \pm  22.32$ & 0.24 &    4.1 \\ 
   3121 & Tamines          & $ 0.27^{+0.48}_{-0.17}$ & $  -7.04 \pm  28.96$ & 0.24 &   26.5 \\ 
   2874 & Jim Young        & $ 0.71^{+0.65}_{-0.71}$ & $   6.02 \pm  26.51$ & 0.23 &    8.5 \\ 
  59150 & 1998\,XV90       & $-2.06^{+0.44}_{-0.28}$ & $  34.49 \pm 154.86$ & 0.22 &   16.7 \\ 
   1772 & Gagarin          & $ 0.05^{+0.02}_{-0.01}$ & $   7.05 \pm  33.65$ & 0.21 &  151.0 \\ 
  51291 & 2000\,KK29       & $-0.00^{+0.00}_{-0.35}$ & $ -43.63 \pm 220.04$ & 0.20 &  125.0 \\ 
  33974 & 2000\,ND17       & $-2.24^{+1.27}_{-0.59}$ & $ -15.20 \pm  84.81$ & 0.18 &    6.8 \\ 
   9582 & 1990\,EL7        & $-2.00^{+2.00}_{-0.48}$ & $  -5.63 \pm  31.94$ & 0.18 &    2.8 \\ 
  38950 & 2000\,ST295      & $-2.39^{+0.79}_{-0.48}$ & $  -8.68 \pm  51.15$ & 0.17 &    3.6 \\ 
  12926 & Brianmason       & $ 2.06^{+0.37}_{-0.52}$ & $  10.77 \pm  63.74$ & 0.17 &    5.2 \\ 
  79056 & 1132\,T-3        & $ 4.10^{+0.46}_{-1.52}$ & $   5.07 \pm  33.43$ & 0.15 &    1.2 \\ 
  33181 & Aalokpatwa       & $-2.30^{+0.54}_{-0.29}$ & $  11.74 \pm  75.86$ & 0.15 &    5.1 \\ 
  38650 & 2000\,ON17       & $ 1.70^{+0.54}_{-0.57}$ & $  10.32 \pm  78.67$ & 0.13 &    6.1 \\ 
   4088 & Baggesen         & $ 0.61^{+0.19}_{-0.27}$ & $  -4.75 \pm  46.27$ & 0.10 &    7.8 \\ 
   9274 & Amylovell        & $-2.63^{+1.28}_{-1.14}$ & $   4.35 \pm  50.55$ & 0.09 &    1.7 \\ 
   6755 & Solov'yanenko    & $ 1.17^{+0.22}_{-0.69}$ & $   4.21 \pm  46.12$ & 0.09 &    3.6 \\ 
  41709 & 2000\,UH56       & $ 1.21^{+0.79}_{-0.51}$ & $   6.62 \pm  82.00$ & 0.08 &    5.5 \\ 
  34290 & 2000\,QQ150      & $ 0.02^{+0.07}_{-0.02}$ & $  -8.56 \pm 104.15$ & 0.08 &  390.8 \\ 
  10406 & 1997\,WZ29       & $-1.35^{+0.32}_{-0.17}$ & $  -8.67 \pm 113.02$ & 0.08 &    6.4 \\ 
   5958 & Barrande         & $-0.05^{+0.02}_{-0.30}$ & $   3.10 \pm  36.51$ & 0.08 &   57.5 \\ 
  49088 & 1998\,RS68       & $-1.99^{+0.42}_{-0.22}$ & $   3.24 \pm  57.54$ & 0.06 &    1.6 \\ 
  20498 & 1999\,RT1        & $ 0.00^{+1.08}_{-0.00}$ & $   2.96 \pm  57.18$ & 0.05 &    2.7 \\ 
   7196 & Baroni           & $-2.17^{+1.15}_{-0.54}$ & $  -1.44 \pm  28.86$ & 0.05 &    0.7 \\ 
  18591 & 1997\,YT11       & $ 0.48^{+0.72}_{-0.34}$ & $  -3.20 \pm  74.72$ & 0.04 &    6.7 \\ 
  17431 & Sainte-Colombe   & $ 2.12^{+0.38}_{-0.53}$ & $  -1.33 \pm  58.18$ & 0.02 &    0.6 \\ 
  32103 & Re'emsari        & $-1.55^{+1.22}_{-0.88}$ & $  -0.27 \pm  65.95$ & 0.00 &    0.2 \\ 
\vspace{-0.3cm}
\enddata
\tablecomments{Solutions for our 134 MBAs, one Mars-crosser (2204 Lyyli), and two NEAs (1685 Toro and 1865 Cerberus). For a detection to be valid, we require both S/N $> 3$ and $\mathcal{S} \leq 2$, where $\mathcal{S}$ is the absolute value of the ratio of the derived and predicted $A_2$. In cases where the predicted $A_2$ was nominally zero, we used the larger error bar value for this calculation instead. Only the NEAs can be considered valid Yarkovsky detections. The remaining MBAs had $A_2$ values associated with very low S/N which were often orders of magnitude larger than expected.}
\end{deluxetable*}
\clearpage

\startlongtable
\begin{deluxetable*}{rlccrcc}
\tablecolumns{7}
\tablecaption{Minimum Arc Lengths Needed for Yarkovsky Detection\label{tab.minarc2}} 
\tablehead{\twocolhead{Designation} & \colhead{$a$} & \colhead{$e$} & \colhead{Predicted $A_2$} & \colhead{Min. $L$} & \colhead{Current $L$} \\
           \twocolhead{} & \colhead{(au)} & \colhead{} & \colhead{($10^{15}$ au day$^{-2}$)} & \colhead{(yr)} & \colhead{(yr)}}
\startdata
  767 & Bondia & 3.119 & 0.1829 & $0.07^{+0.01}_{-0.02}$ & 751--888 & 120 \\ 
  769 & Tatjana & 3.166 & 0.1870 & $0.16^{+0.03}_{-0.16}$ & 532\texttt{+} & 109 \\ 
  963 & Iduberga & 2.248 & 0.1378 & $0.93^{+0.10}_{-0.26}$ & 168--200 & 111 \\ 
 1169 & Alwine & 2.319 & 0.1549 & $-0.91^{+0.64}_{-0.54}$ & 152--299 & 92 \\ 
 1415 & Malautra & 2.223 & 0.0874 & $0.25^{+0.10}_{-0.16}$ & 255--441 & 117 \\ 
 1430 & Somalia & 2.561 & 0.1975 & $0.70^{+0.16}_{-0.70}$ & 213\texttt{+} & 93 \\ 
 1518 & Rovaniemi & 2.226 & 0.1425 & $0.91^{+0.16}_{-0.37}$ & 164--217 & 94 \\ 
 1652 & Herge & 2.251 & 0.1503 & $0.67^{+0.02}_{-0.04}$ & 197--205 & 89 \\ 
 1704 & Wachmann & 2.222 & 0.0873 & $0.94^{+0.14}_{-0.20}$ & 163--190 & 99 \\ 
 1772 & Gagarin & 2.530 & 0.1037 & $0.05^{+0.02}_{-0.01}$ & 583--731 & 83 \\ 
 1773 & Rumpelstilz & 2.437 & 0.1265 & $0.77^{+0.05}_{-0.09}$ & 203--218 & 93 \\ 
 2204 & Lyyli & 2.591 & 0.4049 & $-0.46^{+0.10}_{-0.02}$ & 271--306 & 79 \\ 
 2493 & Elmer & 2.791 & 0.1691 & $-0.77^{+0.69}_{-0.23}$ & 225--615 & 68 \\ 
 2818 & Juvenalis & 2.378 & 0.1495 & $-2.22^{+2.22}_{-0.62}$ & 119\texttt{+} & 62 \\ 
 2840 & Kallavesi & 2.398 & 0.0932 & $0.00^{+0.41}_{-0.00}$ & 264\texttt{+} & 81 \\ 
 2874 & Jim Young & 2.244 & 0.1344 & $0.71^{+0.65}_{-0.71}$ & 150\texttt{+} & 69 \\ 
 3121 & Tamines & 2.229 & 0.0849 & $0.27^{+0.48}_{-0.17}$ & 189--433 & 75 \\ 
 3376 & Armandhammer & 2.348 & 0.0672 & $0.11^{+0.04}_{-0.02}$ & 382--474 & 71 \\ 
 3383 & Koyama & 2.565 & 0.0456 & $0.81^{+0.18}_{-0.25}$ & 201--252 & 72 \\ 
 3510 & Veeder & 2.545 & 0.1288 & $0.71^{+0.06}_{-0.13}$ & 220--246 & 69 \\ 
 3556 & Lixiaohua & 3.169 & 0.2132 & $-0.00^{+0.00}_{-0.51}$ & 359\texttt{+} & 58 \\ 
 3722 & Urata & 2.236 & 0.1994 & $-0.03^{+0.01}_{-0.16}$ & 332--797 & 95 \\ 
 4005 & Dyagilev & 2.452 & 0.1490 & $0.28^{+0.11}_{-0.08}$ & 274--358 & 73 \\ 
 4088 & Baggesen & 2.445 & 0.0585 & $0.61^{+0.19}_{-0.27}$ & 206--291 & 50 \\ 
 4113 & Rascana & 2.260 & 0.0966 & $-0.19^{+0.09}_{-0.48}$ & 201--431 & 58 \\ 
 4285 & Hulkower & 2.643 & 0.1613 & $-1.28^{+1.28}_{-0.24}$ & 176\texttt{+} & 53 \\ 
 4323 & Hortulus & 2.246 & 0.2030 & $1.31^{+0.34}_{-0.56}$ & 139--191 & 71 \\ 
 4399 & Ashizuri & 2.575 & 0.1720 & $-0.30^{+0.17}_{-0.57}$ & 213--466 & 71 \\ 
 4515 & Khrennikov & 2.415 & 0.1531 & $-1.60^{+1.03}_{-0.70}$ & 133--233 & 70 \\ 
 4912 & Emilhaury & 2.302 & 0.1384 & $-1.46^{+1.14}_{-0.25}$ & 141--276 & 69 \\ 
 4928 & Vermeer & 2.147 & 0.1894 & $-1.99^{+0.64}_{-0.43}$ & 113--143 & 40 \\ 
 4959 & Niinoama & 3.149 & 0.0105 & $-0.32^{+0.05}_{-0.04}$ & 410--461 & 72 \\ 
 5589 & De Meis & 2.752 & 0.0406 & $-1.15^{+0.96}_{-0.49}$ & 181--434 & 67 \\ 
 5635 & Cole & 2.385 & 0.2689 & $0.32^{+1.73}_{-0.14}$ & 137--356 & 42 \\ 
 5889 & Mickiewicz & 3.046 & 0.1567 & $-0.70^{+0.15}_{-0.10}$ & 281--327 & 44 \\ 
 5958 & Barrande & 2.348 & 0.1292 & $-0.05^{+0.02}_{-0.30}$ & 273--728 & 67 \\ 
 6499 & Michiko & 2.763 & 0.1412 & $-0.74^{+0.49}_{-0.43}$ & 208--390 & 71 \\ 
 6581 & Sobers & 2.299 & 0.1193 & $-1.99^{+1.21}_{-0.35}$ & 124--193 & 41 \\ 
 6607 & Matsushima & 2.623 & 0.1112 & $-0.71^{+0.71}_{-0.15}$ & 219\texttt{+} & 48 \\ 
 6714 & Montreal & 2.554 & 0.1354 & $0.05^{+0.51}_{-0.02}$ & 251--769 & 43 \\ 
 6755 & Solov'yanenko & 2.442 & 0.0414 & $1.17^{+0.22}_{-0.69}$ & 165--254 & 50 \\ 
 6915 & (1992\,HH) & 2.603 & 0.1244 & $-0.51^{+0.25}_{-0.35}$ & 217--351 & 44 \\ 
 7196 & Baroni & 2.324 & 0.1901 & $-2.17^{+1.15}_{-0.54}$ & 118--176 & 46 \\ 
 7517 & Alisondoane & 2.446 & 0.2632 & $1.04^{+0.07}_{-0.22}$ & 181--205 & 84 \\ 
 7684 & Marioferrero & 2.797 & 0.0534 & $1.06^{+0.27}_{-0.47}$ & 202--281 & 68 \\ 
 8359 & (1989\,WD) & 2.349 & 0.0666 & $-1.85^{+0.67}_{-0.79}$ & 121--168 & 33 \\ 
 9008 & Bohsternberk & 2.177 & 0.1059 & $-1.73^{+0.40}_{-0.20}$ & 126--146 & 39 \\ 
 9158 & Plate & 2.299 & 0.1508 & $-1.55^{+0.76}_{-0.44}$ & 132--192 & 41 \\ 
 9173 & Viola Castello & 2.794 & 0.1175 & $-1.15^{+1.15}_{-0.24}$ & 197\texttt{+} & 47 \\ 
 9234 & Matsumototaku & 2.202 & 0.0992 & $-2.67^{+2.56}_{-0.74}$ & 102--415 & 52 \\ 
 9274 & Amylovell & 2.631 & 0.1576 & $-2.63^{+1.28}_{-1.14}$ & 122--184 & 72 \\ 
 9566 & Rykhlova & 2.361 & 0.2500 & $-1.97^{+0.55}_{-0.23}$ & 131--156 & 46 \\ 
 9582 & (1990\,EL7) & 2.166 & 0.0460 & $-2.00^{+2.00}_{-0.48}$ & 113\texttt{+} & 53 \\ 
10166 & Takarajima & 2.624 & 0.1358 & $-0.00^{+0.00}_{-2.48}$ & 143\texttt{+} & 32 \\ 
10338 & (1991\,RB11) & 3.112 & 0.1555 & $-0.09^{+0.08}_{-0.78}$ & 281--1560 & 47 \\ 
10406 & (1997\,WZ29) & 3.205 & 0.2021 & $-1.35^{+0.32}_{-0.17}$ & 235--275 & 48 \\ 
10456 & Anechka & 2.380 & 0.0423 & $1.82^{+0.36}_{-0.79}$ & 133--180 & 44 \\ 
10656 & Albrecht & 3.176 & 0.0852 & $-0.61^{+0.45}_{-0.14}$ & 307--571 & 69 \\ 
11676 & (1998\,CQ2) & 2.400 & 0.1148 & $-2.64^{+0.34}_{-0.30}$ & 119--132 & 47 \\ 
11823 & Christen & 2.370 & 0.2482 & $1.77^{+0.31}_{-1.44}$ & 135--283 & 63 \\ 
11889 & (1991\,AH2) & 2.762 & 0.1812 & $0.46^{+0.04}_{-0.30}$ & 293--458 & 32 \\ 
12097 & (1998\,HG121) & 2.397 & 0.1456 & $1.04^{+0.44}_{-0.40}$ & 157--219 & 38 \\ 
12374 & Rakhat & 2.552 & 0.3062 & $1.72^{+0.36}_{-0.70}$ & 148--197 & 68 \\ 
12555 & (1998\,QP47) & 2.889 & 0.0809 & $-0.58^{+0.58}_{-0.51}$ & 228\texttt{+} & 48 \\ 
12617 & Angelusilesius & 2.641 & 0.1227 & $-0.00^{+0.00}_{-0.99}$ & 209\texttt{+} & 62 \\ 
12690 & Kochimiraikagaku & 3.005 & 0.1186 & $-0.21^{+0.21}_{-0.14}$ & 381\texttt{+} & 42 \\ 
12705 & (1990\,TJ) & 2.532 & 0.0660 & $-0.51^{+0.31}_{-0.41}$ & 204--379 & 44 \\ 
12883 & (1998\,QY32) & 2.275 & 0.0927 & $-0.00^{+0.00}_{-0.64}$ & 207\texttt{+} & 43 \\ 
12926 & Brianmason & 2.688 & 0.2213 & $2.06^{+0.37}_{-0.52}$ & 149--179 & 71 \\ 
13007 & (1984\,AU) & 2.532 & 0.1339 & $-0.65^{+0.65}_{-0.67}$ & 177\texttt{+} & 116 \\ 
13019 & (1988\,NW) & 2.637 & 0.1766 & $0.71^{+0.35}_{-0.42}$ & 203--341 & 51 \\ 
13059 & Ducuroir & 2.590 & 0.0935 & $1.89^{+0.34}_{-0.60}$ & 147--183 & 40 \\ 
13446 & Almarkim & 3.068 & 0.0961 & $0.00^{+0.07}_{-0.00}$ & 774\texttt{+} & 62 \\ 
13883 & (7066\,P-L) & 3.034 & 0.1517 & $1.01^{+0.09}_{-0.16}$ & 245--272 & 62 \\ 
13925 & (1986\,QS3) & 3.009 & 0.0619 & $-0.57^{+0.16}_{-0.14}$ & 288--360 & 36 \\ 
13936 & (1989\,HC) & 3.205 & 0.0055 & $0.95^{+0.07}_{-0.25}$ & 275--321 & 34 \\ 
13948 & (1990\,QB6) & 2.404 & 0.2317 & $0.43^{+0.63}_{-0.24}$ & 180--362 & 43 \\ 
14031 & Rozyo & 2.588 & 0.1983 & $-0.88^{+0.88}_{-0.46}$ & 180\texttt{+} & 28 \\ 
14257 & (2000\,AR97) & 2.226 & 0.1391 & $0.62^{+0.94}_{-0.62}$ & 141\texttt{+} & 80 \\ 
15920 & (1997\,UB25) & 2.203 & 0.2172 & $-2.40^{+1.72}_{-0.49}$ & 108--194 & 38 \\ 
17431 & Sainte-Colombe & 2.346 & 0.1267 & $2.12^{+0.38}_{-0.53}$ & 124--149 & 33 \\ 
18591 & (1997\,YT11) & 2.621 & 0.0821 & $0.48^{+0.72}_{-0.34}$ & 192--455 & 29 \\ 
18997 & Mizrahi & 2.565 & 0.0872 & $-2.66^{+2.66}_{-0.46}$ & 127\texttt{+} & 36 \\ 
19136 & Strassmann & 2.597 & 0.1050 & $-2.27^{+2.27}_{-0.48}$ & 136\texttt{+} & 33 \\ 
19876 & (7637\,P-L) & 2.348 & 0.1762 & $2.09^{+0.32}_{-1.89}$ & 126--345 & 64 \\ 
20179 & (1996\,XX3)1 & 2.377 & 0.1448 & $-2.36^{+1.30}_{-0.58}$ & 118--177 & 29 \\ 
20498 & (1999\,RT1) & 2.597 & 0.1778 & $0.00^{+1.08}_{-0.00}$ & 198\texttt{+} & 27 \\ 
20515 & (1999\,RO34) & 3.056 & 0.0359 & $0.91^{+0.29}_{-0.39}$ & 239--336 & 37 \\ 
20557 & Davidkulka & 2.378 & 0.1029 & $1.42^{+0.29}_{-0.60}$ & 147--197 & 41 \\ 
20681 & (1999\,VH10) & 2.801 & 0.0899 & $0.99^{+0.27}_{-0.56}$ & 206--320 & 43 \\ 
21041 & (1990\,QO1) & 3.001 & 0.0506 & $-2.06^{+1.29}_{-0.39}$ & 175--278 & 46 \\ 
21842 & (1999\,TH102) & 3.096 & 0.1936 & $-0.44^{+0.44}_{-0.25}$ & 307\texttt{+} & 35 \\ 
22092 & (2000\,AQ199) & 2.554 & 0.1037 & $-0.00^{+0.00}_{-0.23}$ & 360\texttt{+} & 63 \\ 
24181 & (1999\,XN8) & 2.585 & 0.1173 & $-1.38^{+1.38}_{-0.55}$ & 155\texttt{+} & 37 \\ 
25887 & (2000\,SU308) & 2.941 & 0.2040 & $2.34^{+0.32}_{-0.86}$ & 164--208 & 43 \\ 
26176 & (1996\,GD2) & 2.743 & 0.0342 & $-2.22^{+0.59}_{-0.58}$ & 145--180 & 27 \\ 
26520 & (2000\,CQ75) & 2.679 & 0.2641 & $1.04^{+0.34}_{-1.04}$ & 187\texttt{+} & 71 \\ 
27259 & (1999\,XS136) & 2.570 & 0.1297 & $1.18^{+0.11}_{-0.14}$ & 181--197 & 71 \\ 
28709 & (2000\,GY96) & 3.130 & 0.1585 & $-1.41^{+0.28}_{-0.15}$ & 224--255 & 28 \\ 
28736 & (2000\,GE133) & 2.282 & 0.0664 & $-2.09^{+0.49}_{-0.18}$ & 125--143 & 30 \\ 
30072 & (2000\,EP93) & 2.689 & 0.1877 & $0.90^{+0.26}_{-0.54}$ & 201--322 & 24 \\ 
31755 & (1999\,JA96) & 2.938 & 0.0705 & $-0.95^{+0.95}_{-0.92}$ & 188\texttt{+} & 25 \\ 
31829 & (1999\,XT12) & 2.993 & 0.0288 & $-0.59^{+0.59}_{-0.89}$ & 213\texttt{+} & 31 \\ 
32103 & Re'emsari & 2.357 & 0.2537 & $-1.55^{+1.22}_{-0.88}$ & 126--280 & 32 \\ 
32507 & (2001\,LR15) & 2.789 & 0.2119 & $-2.88^{+0.71}_{-0.56}$ & 137--165 & 32 \\ 
32776 & Nriag & 2.608 & 0.1699 & $-1.05^{+0.66}_{-0.92}$ & 156--301 & 36 \\ 
33116 & (1998\,BO12) & 2.335 & 0.1673 & $1.52^{+0.43}_{-0.46}$ & 136--174 & 25 \\ 
33181 & Aalokpatwa & 2.702 & 0.0781 & $-2.30^{+0.54}_{-0.29}$ & 146--171 & 31 \\ 
33974 & (2000\,ND17) & 2.448 & 0.1082 & $-2.24^{+1.27}_{-0.59}$ & 124--191 & 27 \\ 
34290 & (2000\,QQ150) & 2.988 & 0.0914 & $0.02^{+0.07}_{-0.02}$ & 643\texttt{+} & 26 \\ 
35595 & (1998\,HO116) & 2.562 & 0.0835 & $-0.24^{+0.24}_{-0.27}$ & 261--1805 & 24 \\ 
37377 & (2001\,VP46) & 2.580 & 0.1647 & $-0.08^{+0.03}_{-0.05}$ & 447--628 & 71 \\ 
38650 & (2000\,ON17) & 2.633 & 0.1982 & $1.70^{+0.54}_{-0.57}$ & 150--198 & 34 \\ 
38950 & (2000\,ST295) & 2.272 & 0.0901 & $-2.39^{+0.79}_{-0.48}$ & 113--143 & 24 \\ 
40104 & (1998\,QE4) & 2.995 & 0.0838 & $-1.16^{+0.20}_{-0.41}$ & 209--254 & 24 \\ 
40413 & (1999\,RS10) & 2.445 & 0.0671 & $1.64^{+0.51}_{-1.26}$ & 138--276 & 25 \\ 
41709 & (2000\,UH56) & 2.677 & 0.1421 & $1.21^{+0.79}_{-0.51}$ & 161--245 & 47 \\ 
42490 & (1991\,SU) & 2.414 & 0.1384 & $1.80^{+0.22}_{-1.80}$ & 140\texttt{+} & 38 \\ 
44612 & (1999\,RP27) & 2.198 & 0.1783 & $-2.20^{+1.46}_{-2.61}$ & 88--186 & 25 \\ 
45898 & (2000\,XQ49) & 1.956 & 0.0733 & $2.53^{+0.86}_{-1.12}$ & 90--127 & 41 \\ 
47127 & (1999\,CJ103) & 3.100 & 0.1674 & $-0.83^{+0.35}_{-0.29}$ & 252--352 & 46 \\ 
49088 & (1998\,RS68) & 2.314 & 0.1011 & $-1.99^{+0.42}_{-0.22}$ & 128--147 & 41 \\ 
50219 & (2000\,AL237) & 2.569 & 0.0976 & $-0.16^{+0.05}_{-0.76}$ & 208--496 & 35 \\ 
50776 & (2000\,FS12) & 2.651 & 0.1774 & $0.75^{+0.49}_{-0.44}$ & 192--334 & 30 \\ 
51291 & (2000\,KK29) & 3.185 & 0.1514 & $-0.00^{+0.00}_{-0.35}$ & 421\texttt{+} & 26 \\ 
51822 & (2001\,OB25) & 3.170 & 0.0598 & $-0.26^{+0.16}_{-1.63}$ & 212--671 & 31 \\ 
56131 & (1999\,CY48) & 2.368 & 0.0932 & $1.05^{+1.96}_{-1.05}$ & 116\texttt{+} & 32 \\ 
57429 & (2001\,SX33) & 2.777 & 0.0647 & $-1.94^{+1.94}_{-0.56}$ & 155\texttt{+} & 24 \\ 
59150 & (1998\,XV90) & 3.184 & 0.1196 & $-2.06^{+0.44}_{-0.28}$ & 196--227 & 38 \\ 
66003 & (1998\,OX6) & 2.860 & 0.0647 & $-0.17^{+0.07}_{-0.96}$ & 222--586 & 24 \\ 
79056 & (1132\,T-3) & 2.374 & 0.2508 & $4.10^{+0.46}_{-1.52}$ & 99--124 & 44 \\ 
87932 & (2000\,SW343) & 2.606 & 0.1249 & $1.05^{+0.17}_{-0.38}$ & 189--241 & 22 \\ 
89932 & (2002\,EV85) & 2.684 & 0.1969 & $2.14^{+0.57}_{-1.87}$ & 143--359 & 27 \\ 
92930 & (2000\,RH26) & 2.632 & 0.2395 & $-0.03^{+0.01}_{-0.06}$ & 534--1037 & 22 \\ 
99028 & (2001\,DC98) & 2.631 & 0.2023 & $-3.79^{+2.39}_{-0.65}$ & 114--181 & 23 \\ 
\vspace{-0.3cm}
\enddata
\tablecomments{Minimum observational arc lengths $L$ needed to detect a predicted $A_2$ value at S/N = 3 for our sample of 134 observed MBAs and one Mars-crosser (2204 Lyyli), calculated as described in \S\ref{sec.minarc}. The semimajor axis $a$ and eccentricity $e$ are taken from the JPL Small-Body Database. The predicted $A_2$ values are estimated from TPM-derived parameters as described in \S\ref{sec.tpm}. The current arc lengths are the total time span of the existing observations for each asteroid, last retrieved on 2023 January 1. The minimum arc length spans cover the $1\sigma$ range in the predicted $A_2$ values. In instances where the predicted $A_2$ reaches zero, the minimum arc length is essentially infinite; thus, we only report a lower bound for such asteroids.}
\end{deluxetable*}

\end{document}